\documentclass[amssymb,prb,preprint,aps]{revtex4}
\usepackage{epsfig}
 \evensidemargin
 \oddsidemargin
\begin{document}
\sloppy\baselineskip 15pt
\title{Double exchange mechanisms for Mn doped III-V
ferromagnetic semiconductors}
\author{P. M. Krstaji\'{c},  F. M. Peeters}
\affiliation{Departement Natuurkunde, Universiteit Antwerpen
(Campus Drie Eiken), Universiteitsplein 1, B-2610 Antwerpen,
Belgium}\author{V. A. Ivanov\cite{perm}}
\affiliation{N.S.
Kurnakov Institute of the General and Inorganic Chemistry of the
Russian Academy of Sciences, Leninskii prospect 31, 117 907
Moscow, Russia}
\author{V. Fleurov\cite{aut2}}
\affiliation{Raymond and Beverly Sackler Faculty of Exact Sciences,
School of Physics and Astronomy, Tel Aviv University, Ramat Aviv,
69 978 Tel Aviv, Israel}
\author{K. Kikoin}
\affiliation{Physics Department, Ben-Gurion University, 84 105
Beer-Sheva, Israel}
\date{\today}

\begin{abstract}
A microscopic model of indirect exchange interaction between
transition metal impurities in dilute magnetic semiconductors
(DMS) is proposed. The hybridization of the impurity d-electrons
with the heavy hole band states is largely responsible for the
transfer of electrons between the impurities, whereas Hund rule
for the electron occupation of the impurity d-shells makes the
transfer spin selective. The model is applied to such systems as
$n-$type GaN:Mn and $p-$type (Ga,Mn)As, $p-$type (Ga,Mn)P. In
$n-$type DMS with Mn$^{2+/3+}$ impurities the exchange mechanisms
is rather close to the kinematic exchange proposed by Zener for
mixed-valence Mn ions. In $p-$type DMS ferromagnetism is governed
by the kinematic mechanism involving the kinetic energy gain of
heavy hole carriers caused by their hybridization with 3d
electrons of Mn$^{2+}$ impurities. Using the molecular field
approximation the Curie temperatures $T_C$ are calculated for
several systems as functions of the impurity and hole
concentrations. Comparison with the available experimental data
shows a good agreement.
\end{abstract}
\maketitle

\section{Introduction}

Dilute magnetic semiconductors (DMS) are semiconductors, in which
transition or rare-earth metal atoms randomly replace a fraction
of atoms in one sublattice. Transition metal (TM) impurities, due
to their high abundance and diffusivity, can easily enter the host
semiconductor. In the group IV elemental semiconductors, they
occupy mainly interstitial positions, whereas in III-V
semiconductors the 3d TM impurities usually substitute group III
atoms. The metastability of the zinc blende phase of the DMS
(III,Mn)V compounds and low solubility of manganese in these
materials were the major obstacles for a synthesis of dilute
magnetic semiconductors.\cite{Hayashi} However, the idea to
combine in DMS the charge degrees of freedom of hole or electron
carriers with the spin degrees of freedom of magnetic impurities
became reality after a new doping technique based on
non-equilibrium low temperature molecular beam epitaxy
(LT-MBE)\cite{fmatsukara,kikkawa} was developed. The
discovery\cite{Ohno0} of ferromagnetic ordering in manganese doped
InAs with the $T_C=7.5$K Curie temperature fuelled DMS studies
that resulted in fabrication of (Ga,Mn)As compounds with $T_C =
110$K \cite{fmatsukara,OhnoS} or even $T_C = 140$K.\cite{Gall2}
More recently\cite{kw02,kp03,g03} Curie temperatures exceeding
150K were reported in (Ga,Mn)As. An above room temperature
ferromagnetism was announced\cite{sonoda,reed} also in GaN and
{\em{p}}-type GaP doped with Mn. Advanced III-V growth technique
such as metal organic vapor phase epitaxy (MOVPE), or metal
organic chemical vapor deposition (MOCVD) together with LT-MBE can
produce good quality DMS with various element combinations.

A unique coexistence of high-temperature magnetism and
semiconductor properties opens new venues of fundamental studies
and applications of DMS's. The DMS's may be quite promising in
spintronics, information technology,\cite{kikkawa,crooker} quantum
computing,\cite{Vinc} in manipulation of magnetism by an electric
field\cite{ohno-field} and/or by illumination,\cite{oiwa} in
devices governed by the giant magnetoresistance
effect,\cite{Tanaka} where the interlayer tunneling resistance is
changed under the action of a magnetic field. Among the variety of
applications one cannot exclude other functionalities such as
nanostructure aspect of DMS: \cite{konig} the DMS based quantum
wells are expected to have unusual ferromagnetic properties with
$T_C$ (as shown in field effect transistors\cite{ohno-nature}) and
high coercitivities\cite{kawakami} tunable electrically, rather
than magnetically.\cite{BLee}

Although ferromagnetic order in (Ga,Mn)As has been observed for
the first time as early as\cite{Shen} 1996 with $T_C=60$K, its
microscopic origin from the theoretical point of view still
remains not well understood. Extrapolating from (Ga,Mn)As, FM
ordering with high $T_C$ was predicted for $p$-type (Ga,Mn)N
\cite{Domcf} and found in Ref. \onlinecite{reed}. However, a
direct extrapolation of the trends known for the GaAs-based
material characterized by a relatively narrow forbidden energy gap
to the wide gap GaN DMS is not well founded. The differences in
formation of the ferromagnetic order arise from the differences in
the structure of chemical bonds between the Mn impurity and
valence electrons in various III-V host semiconductors. The
purpose of the present paper is to pinpoint these differences and
to construct a general microscopic theory of the double exchange
(superexchange) in dilute magnetic semiconductors.

There are three approaches, which one may follow to study magnetic
states of DMS's. According to the first approach, one chooses an
effective spin Hamiltonian and calculates the corresponding
exchange coupling constant in the course of this derivation. The
major part of the available descriptions of the FM order in such
DMS's as (Ga,Mn)As, (Ga,Mn)P and (Ga,Mn)N, are based on
semi-phenomenological models, which postulate the existence of
local magnetic moments on the Mn sites, as well as an indirect
exchange between these moments and the electrons in the valence
band of the host
semiconductor.\cite{dietl2,dietl3,Semenov,konig,Bhatt} Some of
these theories emphasize the role of shallow acceptor levels
existing in the first two of these systems \cite{Duga} or of
resonance levels.\cite{Inoue}

The second approach is based on an extended cluster method, where
a cubic supercell with a magnetic Mn ion in its center is used to
calculate the density of spin polarized states of a "homogeneously
doped" crystal (see e.g., Refs. \onlinecite{sanvito,Sato}). The
resulting picture provides information about positions and
occupation of Mn-related majority and minority "bands". Sometimes
an effective $s-d$ exchange Hamiltonian is introduced
phenomenologically in this picture, and the effective exchange
constant is chosen to fit the numerical data (see Ref.
\onlinecite{sanvito} and the first paper in Ref.
\onlinecite{Sato}).

According to the third approach, realistic exchange parameters
should be derived within the framework of a microscopic theory
based on the knowledge of the electronic structure of III-V
semiconductors doped with TM impurities. An exhaustive microscopic
theory for isolated TM impurities in semiconductors was
constructed more than two decades ago starting with the papers of
Refs. \onlinecite{Fleurov1,Haldane} (see monograph Ref.
\onlinecite{Kikoin} for a detailed description and Ref.
\onlinecite{Zunger} where a powerful numeric approach was
developed). This theory contains all necessary ingredients for an
accurate derivation of the indirect magnetic exchange between
interacting impurities. Our first attempt\cite{KIP} to apply this
theory to (Ga,Mn)As resulted in a quantitative description of the
hole concentration dependent Curie temperature $T_C$ in
{\em{p}}-type DMS's within the framework of a very simple model
Hamiltonian, in which the effective {\em kinematic exchange}
results from hopping between the occupied $d$-states of the
half-filled $d$-shell of the Mn impurity via empty heavy hole
({\em hh}) states near the top of the valence band.

Here we present a general theory of the double exchange
interaction in Mn-doped III-V semiconductors, which covers both
$p$- and $n$-type materials. The theory establishes differences
between the exchange mechanisms in these two cases and
demonstrates that the conventional Zener double exchange mechanism
\cite{Zener} is realized only in $n$-type DMS (presumably, in
(Ga,Mn)N).

It is well known that the TM ions may introduce impurity levels in
the forbidden energy gap. Among them the manganese impurity in
GaAs and some other semiconductors exhibit peculiar features.
According to its position in the series of the TM elements, a
neutral impurity Mn$^{3+}$ in the Ga-site is expected to have the
$3d^4$ configuration. However, the Mn ion retains its fifth
electron in its $3d$ shell, because of an exceptional stability of
the high-spin half-filled $3d^5$ state resulting from the strong
intra-atomic Hund interaction. The manganese impurities not only
introduce magnetic moments in the {\em{p}}-type (III,Mn)V
compounds but also create potentials attracting holes, i.e. act as
acceptors. Therefore, the neutral substitution impurity state is
$A^0(3d^{5}\bar{p})$, where $\bar{p}$ stands for the loosely bound
hole, and the manganese provides both holes and magnetic moments
to the host, $A$ is the spectroscopic notation indicating the
irreducible representation describing this state (more about these
notations can be found, e.g. in Ref. \onlinecite{Kikoin})

Existence of the complex Mn $d^5\bar{p}$ in lightly doped bulk
GaAs is detected by electron spin resonance measurements
\cite{schneider} and by infrared spectroscopy.\cite{Linnarson}
These studies discovered an acceptor level inside the energy gap
around 110 meV above the top of the valence band. This value is
substantially higher than 30 meV resulting from conventional
effective mass theory,\cite{schneider} which already implies that
treatment of the manganese impurities in III-V compounds requires
a more refined approach. On the other hand, the presence of
negatively ionized complexes $A^{-}(3d^{5})$ is detected in
(Ga,Mn)As epilayers grown by low temperature molecular beam
epitaxy,\cite{szczytko} which indicates that both mobile and
localized holes exist in ferromagnetic (Ga,Mn)As. The coexistence
of weakly and strongly localized states is also in accordance with
{\em ac} conductivity experiments. \cite{Nagai}

Apparently, a similar impurity electronic structure is realized in
(Ga,Mn)P. But in contrast, Mn in the wide gap GaN semiconductor
releases one of its $d$-electrons to the valence band, which is
typical of all other TM impurities, and remains in the
Mn$^{3+}(d^4)$ state, unless the sample is deliberately $n$-doped.
The latter state is neutral, so there is no need in binding a hole
to maintain the neutrality of the ground state. This leaves no
room for a direct extrapolation from (Ga,Mn)As to (Ga,Mn)N, and
the kinematic inter-impurity exchange mechanisms in these two
systems should be considered separately.

In this paper we will develop a general theory of ferromagnetic
exchange in DMS's. It turns out that the Zener mechanism
\cite{Zener} in its classical form is realized in $n$-type
(Ga,Mn)N, whereas the $p$-type systems (Ga,Mn)As, (Ga,Mn)P and,
possibly, $p$-(Ga,Mn)N are examples of another type of indirect
(kinematic) exchange interaction, which was not observed in the
family of Mn oxides described by the Zener theory of ferromagnetic
insulators. The paper is organized as follows. Section II presents
a detailed description of the microscopic model including the
calculation of the exchange interaction between TM impurities.
Since this calculation appears to be rather involved, its
mathematical details are given in the Appendix. In Section III we
apply the results obtained in Section II to {\em{p}}-type
semiconductors, find the kinematic exchange interaction and using
the molecular field approximation calculate the Curie temperature
as a function of the Mn content and hole concentration for
{\em{p}}-type (Ga,Mn)As and (Ga,Mn)P. The results are compared
with available experimental data. The $n$-type (Ga,Mn)N DMS's,
which possesses specific features differing it from other DMS's,
is discussed in Section IV. Finally, the scope of the theory and
comparison with other approaches are summarized in the concluding
Section V.

\section{Indirect exchange interaction between two magnetic impurities}

According to the general principles of the theory of isolated TM
in semiconductors,\cite{Kikoin,Zunger} the electronic spectrum of
these impurities in III-V semiconductors is predetermined by the
structure of chemical bonds (hybridization) between the
$3d$-orbitals of impurity ions and the $p$-orbitals of the valence
band electrons, whereas their magnetic state is governed by the
Coulomb and exchange interactions within the $3d$-atomic shell
(intra-atomic electron-electron correlations), modified by
hybridization with the electron states of the crystalline
environment.

The localized TM impurity $d$-states possess $t_2$ and $e$
symmetry of the crystal point symmetry group $T_d$ as revealed in
the theoretical studies of electronic properties of isolated TM
impurities in semiconductors carried out in the 70's and 80's
(see, e.g. Refs.
\onlinecite{Kikoin77,Hemstreet,Picoli,Singh,Fleurov2}). The $e$
states are practically nonbonding $d$-states of the impurity 3$d$
shell, whereas the $t_2$ states form bonding and antibonding
configurations with the $p$-states of the valence band. The two
latter types of states are classified as the crystal field
resonances (CFR), in which the $d$-component dominates, and the
dangling bond hybrids (DBH) with predominantly $p$-type
contribution of the valence band states. Crucially important is
that the absolute positions of CFR levels relatively weakly depend
on the band structure of the host semiconductor material. The CFR
levels are pinned mainly to the vacuum energy, and this pinning is
modulated by the counterbalanced interactions with valence and
conduction band states.\cite{Kikoin2,Kikoin}  The DBH states are
more intimately connected with the peak in the density of states
of the heavy hole band, which dominates the {\em
pd}-hybridization. The variety of electronic and magnetic
properties of Mn and other TM ions in different III-V host
semiconductors stems from these universal chemical trends.

Due to a large mismatch between positions of the valence bands in
GaAs, GaP on the one side and in GaN on the other side, the
Mn($d^{5}/d^{4}$) CFR level falls deep in the valence band in
GaAs, GaP and appears within the forbidden energy gap of GaN. The
CFR nature of Mn $(d^5/d^4)$ $t_2$-level near the middle of the
gap for (Ga,Mn)N is confirmed experimentally\cite{graf} and
follows also from numerical calculations.\cite{Sato} This
$t_2$-level is {\em empty} in the neutral state Mn$^{3+}(d^4)$ of
the substitution impurity. It becomes magnetically active only in
$n$ doped materials, e.g. in (Ga,Mn)N, where part of the Mn
impurities capture donor electrons and transform into charged ions
Mn$^{2+}(d^5)$. A similar behavior is characteristic of all other
light TM ions in all III-V compounds including GaAs and
GaP.\cite{Hemstreet,Zunger,Kikoin}

Since the position of the valence band in GaAs and GaP is
substantially higher in the absolute energy scale (with the vacuum
energy as the reference point) than those in GaN, the CFR $t_2$
level ($d^5/d^4$) in the (Ga,Mn)As and (Ga,Mn)P DMS's falls below
the center of gravity of the valence band. Therefore, this CFR
level is always occupied, Mn ions retain their fifth electron in
the $3d$-shell, and the neutral substitution impurity state is
($3d^5\bar{p}$), where $\bar{p}$ stands for a loosely bound hole
occupying the relatively shallow acceptor level in the forbidden
energy gap of GaAs and GaP semiconductors.

Both of the above mentioned situations may be described within the
resonance scattering model of the TM impurity states in
semiconductors based on the Anderson single impurity Hamiltonian,
proposed originally for TM impurities in metals,\cite{Anderson}
and later modified for semiconductors in Refs.
\onlinecite{Fleurov1,Haldane,Picoli,Kikoin2,Fleurov2}. An
extension of the Anderson model to two TM impurities in metals was
proposed in Ref. \onlinecite{Alex} resulting in a ferromagnetic
coupling in TM doped metals. Using this model Ref
\onlinecite{Caroli} derived an interaction similar to the RKKY
interaction. Here we discuss a model describing interaction
between Mn (and other TM) impurities in a semiconductor host. In
accordance with Hund rule for the electron occupation in Mn ions
in a tetrahedral environment, \cite{Kikoin3,Zunger,Kikoin} two
competing states Mn(d$^4$) and Mn (d$^5$), involved in the double
exchange in DMS, have the $d^4(e^2t^2)$ and $d^5(e^2t^3)$
configurations, respectively. The next charged state $d^6(e^3t^3)$
has a much higher energy due to the strong intra-ionic Coulomb
interaction. Therefore, the 'passive' nonbonding $e$-electrons may
be excluded from our consideration. The indirect exchange between
magnetic moments is mediated by virtual transitions of $t_2$
electrons into unoccupied valence band states.

The Hamiltonian for two TM impurities in a III-V semiconductor has
the form:
\begin{equation}
H=H_h + H_d+H_{hd},  \label{And1}
\end{equation}
where the band Hamiltonian
\begin{equation}
H_h = \sum_{{\bf p},\sigma}\varepsilon_{{\bf p} }^{h}c_{{\bf
p}h\sigma }^{\dagger }c_{{\bf p}h\sigma } \label{And11}
\end{equation}
includes only the heavy hole band. Here $c_{{\bf p}h\sigma
}^\dagger(c_{{\bf p} h\sigma })$ is the creation (annihilation)
operator of a hole with momentum $ {\bf p}$ and spin orientation
$\sigma $ in the {\em hh} band of the semiconductor with the
energy dispersion $\varepsilon _{{\bf p}}^h$. The heavy hole band
gives the dominant contribution to the formation of the impurity
states,\cite{Zunger,Kikoin} and governs the onset of ferromagnetic
order. The second term in the Hamiltonian (\ref {And1}) describes
the electrons within the Mn atoms with a possible account of the
crystal field of the surrounding atoms. In principle, we have to
write down a multielectron Hamiltonian describing the degenerate
states of the $d$ shell, and including Coulomb and exchange
interactions. However, as we show below it is usually sufficient
to consider the non-degenerate version of this Hamiltonian
\begin{equation}
H_d =\sum_{i\sigma } \left(E_d \hat{n}_{i\sigma} + \frac{U}{2}
\hat{n}_{i\sigma} \hat{n}_{i-\sigma} \right), \label{And2}
\end{equation}
which simplifies the calculation considerably. Here $E_d$ is the
atomic energy level of the localized Mn $t_{2}$-electrons and $U$
is the Anderson-Hubbard repulsive parameter; $\hat{n}_{i\sigma} =
d_{i\sigma }^\dagger d_{i\sigma }$ is the occupation operator for
the manganese $t_2$ - electrons on the impurities, labelled
$i=1,2$ in (III,Mn)V DMS. The situations when the degeneracy of
the $t_2$ states is important and the exchange interaction (Hund
rule) in the multielectron atom starts playing an essential part
will be outlined and the appropriate corrections will be
introduced, when necessary.

The last term in Eq. (\ref{And1}) describes the scattering of
heavy holes by the impurities
\begin{equation}\label{And3}
H_{hd}=H_{hd}^{(r)}+H_{hd}^{(p)},
\end{equation}
\begin{eqnarray}
H_{hd}^{(r)} & = & \sum_{{\bf p},~\sigma ,j} \left( V_{{\bf
p}d}c_{{\bf p}h\sigma}^\dagger d_{i\sigma }e^{i{\bf pR_j}/ \hbar}
+h.c.\right) ,
\nonumber \\
H_{hd}^{(p)} &=&\sum_{{\bf pp^{\prime }},~\sigma ,j}W_{{\bf pp}'}
c_{{\bf p}h\sigma }^{\dagger }c_{{\bf p}^{\prime }h\sigma
}e^{i{\bf (p-p')R_j}/ \hbar}. \nonumber
\end{eqnarray}
Here $V_{{\bf p}d}$ is the $p$-$d$ hybridization matrix element,
and $W_{{\bf pp}'}$ is the scattering matrix element due to the
difference of the pseudopotentials of the host and the substituted
atoms. The direct overlaps between the $d$-electron wave functions
of the neighboring impurities in the DMS and the Coulomb
interaction between them are neglected.

As mentioned above, the orbital degeneracy of the $t_2$ states is
not taken into account in this simplified model. In reality, the
orbital quantum numbers are important at least in three respects.
First, the half-filled $d^5$ subshell of Mn is occupied by $e$ and
$t_2$ electrons with parallel spins in accordance with the Hund
rule, so the sixth electron can be captured in the $d^6$
configuration only with the opposite spin. The energy cost of this
capture is $\sim U + J$, where $J\ll U$ is the exchange energy.
This feature of the impurity is preserved in the above simplified
Hamiltonian: (\ref{And2}) the reaction $d^5+e\to d^6$ is changed
for $d^1+e\to d^2$, with the spin of the second electron opposite
to the first one. The energy cost of this reaction is $U$, and the
principal features of the ion, important for the formation of the
localized magnetic moment, are practically the same as in the
original atom.

Second, it is the Hund rule that requires that the total angular
moments of the Mn atoms be parallel to allow the indirect
inter-impurity interaction between the high-spin $d^5$ states via
the {\em hh} valence band states. We will take the Hund rule
explicitly into account, when calculating $T_C$, and it will be
shown below, that the energy gain due to this indirect
interaction, monitored by the Hund rule, leads eventually to
ferromagnetic ordering in DMS's.

Third, the three-fold degeneracy of the $t_2$ electrons is also
manifested in the statistics of the localized states, and
therefore it influences the position of the Fermi energy in the
energy gap. These statistics are described in the Appendix. The
degeneracy introduces also numerical factors in the effective
exchange constants. These factors are also calculated in the
Appendix.

Having all this in mind, we proceed with the derivation of the
indirect inter-impurity exchange within the framework of the
non-degenerate Alexander-Anderson model, whereas the corresponding
corrections for the real orbital structure of the Mn ion will be
made when necessary. To calculate the energy of the exchange
interaction between two magnetic impurities, one should find the
electronic spectrum of the semiconductor in the presence of two
impurities. Each impurity perturbs the host electron spectrum
within a radius $r_b$. Inter-impurity interaction arises, provided
the distance between the impurities $R_{ij}$ is comparable with
$2r_b$. General equations for the two-impurity states and the
corresponding contribution to the exchange energy are derived in
Appendix. Here we present only the final equations, which will be
used in the derivation of the effective exchange coupling and the
Curie temperatures for the specific DMS's.

The quantity, which we actually need here is the impurity related
correction to the energy of the system. It is given by the
standard formula (see, {\em e.g.,} Ref.
\onlinecite{Caroli,Haldane})
\begin{equation}
\label{energy1} E^{magn} = \frac{1}{\pi }{\rm Im}\int_{-\infty
}^\infty \varepsilon \cdot {\rm Tr}\Delta G[\varepsilon - i \delta
\mbox{sign}(\varepsilon - \mu)] d \varepsilon - \sum_i U
n_{di\uparrow} n_{di\downarrow}.
\end{equation}
 $\Delta {\sf G}= \sf{G}(\varepsilon) -
\sf{g}^0(\varepsilon)$ is the two impurity correction to the total
one-particle Green function ${\sf G}\left( \varepsilon \right)
=\left( \varepsilon - {\sf H} \right)^{-1}$

Here ${\sf G}$ is defined as a matrix in $\left( h,d \right)$
space, and $\sf{g}^0$ is the same matrix in the absence of
impurity scattering, see Eq. (\ref{AAA5}). The Green function is
calculated in the self-consistent Hartree-Fock approximation for
the inter-impurity electron-electron repulsion {\em U}, which is
sufficient for a description of magnetic correlations in TM
impurities in semiconductors. This solution is described in the
Appendix. Then the part of the two-impurity energy (\ref{energy1})
responsible for the inter-impurity magnetic interaction may be
found (see Eqs. (\ref{property}) and (\ref{calc3})). As shown in
Refs. \onlinecite{Fleurov1,Fleurov2}, the resonance scattering
alone may result in creation of CFR and DBH levels split off from
the valence band, provided the scattering potential is comparable
with the bandwidth. However, in the case of shallow DBH states the
potential scattering may dominate their formation. Like in the
well-known Koster-Slater impurity model, the potential scattering
in our model is described by a short-range momentum independent
coupling constant, $W_{\bf pp'} \approx W$, and the same
approximation is adopted for the resonance scattering parameter,
$V_{{\bf p}d}\approx V$.

Then (\ref{energy1}) acquires the compact form
\begin{equation}\label{calc6}
\Delta E = - \frac{1}{\pi}{\rm Im}
\int_{\varepsilon_{hb}}^{\varepsilon_F} d\varepsilon \left[ \ln
{\sf R}^{\sigma }(\varepsilon )+ \ln {\sf Q}^{\sigma }(\varepsilon
)\right] + E_{loc}(\varepsilon < \mu).
\end{equation}
Here the first term in the r.h.s. describes the band contribution
to the exchange energy, and the integration is carried out from
the bottom of the $hh$ band $\left( \varepsilon_{hb} \right)$ to
the Fermi level $\varepsilon_F$. In this integral the kinetic
energy gain of the band electrons due to scattering by
ferromagnetically aligned impurities is incorporated. The
contributions of resonance and potential scatterings are given by
the first and second term in this integral, respectively. The last
term $\Delta E_{loc}$ is the contribution of the occupied
localized CFR and/or DBH states. These states are described by
zeros of the functions ${\sf R}^{\sigma }(\varepsilon )$ and ${\sf
Q}^{\sigma }(\varepsilon )$ in the discrete part of the energy
spectrum (see Eqs. (\ref{detB}) and (\ref{detA}), respectively).

In all the cases the energy gain results from the indirect
spin-dependent inter-impurity overlap, and the mechanism of the
effective exchange interaction may be qualified as kinematic
exchange. In general, double exchange favors a FM order, because
the splitting of the energy levels belonging to two adjacent
impurities occurs due to electron hopping via unoccupied band or
impurity-related states {\em without spin reversal}. Were the
impurity angular moments non-parallel, the Hund rule would
suppress the probability of the electron, with spin parallel to
the first impurity angular momentum, to hop onto the second
impurity with a "wrong" direction of the angular momentum.

The level splitting, when not suppressed by the Hund rule, results
in an energy decrease provided that not all of the available band
and impurity levels are occupied. One should note that this
kinematic exchange cannot be reduced to any conventional double
exchange mechanism because of an actual interplay between three
contributions to the magnetic energy, namely, the scattered
valence band electrons, the CFR states and the DBH states. We
start with the analysis of the band contribution $\Delta
E_{b,ex}$. Since in all the cases we deal with nearly filled $hh$
bands, it is convenient to calculate the energy in the hole
representation.

By means of a simple trick (as described in the Appendix) the
contribution of the {\em hh} band in the basic equation
(\ref{calc6}) may be transformed into
\begin{equation}\label{calc2r}
\Delta E_{b,ex} = \frac{1}{\pi}{\rm Im}
\int^{\varepsilon_{ht}}_{\varepsilon_F} d\varepsilon  \ln
\frac{{\sf R}^\sigma (\varepsilon )}{{\sf R}_0^\sigma (\varepsilon
)},
\end{equation}
where $\varepsilon_{ht}$ stands for the top of the valence band.
This equation is obtained from the more general equation
(\ref{calc2}) by neglecting the potential scattering in the band
continuum. Then, inserting Eq. (\ref{detB}) for the matrix ${\sf
R}^{\sigma }(\varepsilon )$ we obtain
\begin{equation}\label{calc4}
\Delta E_{b,ex} = -\frac{1}{2\pi}\mbox{Im}
\int^{\varepsilon_{ht}}_{\varepsilon_F}d\varepsilon
\ln\left\{\frac{[g_d^{-1}(\varepsilon - i\delta) - V^2
L_{11}(\varepsilon - i\delta)]^2 - V^4 L_{12}^{2}(\varepsilon -
i\delta)}{g_d^{-1}(\varepsilon - i\delta) - V^2 L_{11}(\varepsilon
- i\delta)}\right\}.
\end{equation}
Here
\begin{equation}\label{green}
L_{ij}(\varepsilon - i\delta)\equiv P_{ij}(\varepsilon) +
\frac{i}{2} \Gamma_{ij}(\varepsilon)
\end{equation}
are the standard lattice Green functions (\ref{sec1}) of the
continuous argument $\varepsilon$. Retaining only the leading
(quadratic) terms in $L_{12}$ in Eq. (\ref{calc4}), the exchange
energy due to the {\em hh} band becomes
\begin{eqnarray}
\Delta E_{b,ex} =  -\frac{V^4}{\pi}
\int^{\varepsilon_{ht}}_{\varepsilon_F} d\varepsilon
\frac{P_{12}(\varepsilon)\Gamma_{12}(\varepsilon)}{ [\varepsilon -
E_d - V^2 P_{11}(\varepsilon)]^2 + \displaystyle \frac{V^4}{4}
\Gamma^2_{11} (\varepsilon)}. \label{calc7}
\end{eqnarray}

Eq. (\ref{calc4}) with $L_{12}(\varepsilon) \neq 0$ holds only if
the impurity angular moments are parallel to each other
(ferromagnetic case), since the electron hopping between the
impurities via the {\em hh} band takes place without spin-flips.
If the angular moments are non-parallel, the Hund energy does not
allow the electron to hop to the neighboring impurity with a
"wrong" angular momentum direction. So the impurities with
non-parallel angular moments cannot exchange their electrons,
which effectively means that for them $L_{12} = L_{21}=0$. Thus,
Eq. (\ref{calc7}) is the energy decrease due to the kinematic
indirect exchange between a pair of magnetic impurities with
parallel angular moments. If the moments are not parallel the
corresponding energy decrease is zero.

Another contribution to the magnetic energy $\Delta E$ (Eq.
(\ref{calc6})) originates from the inter-impurity hopping via
empty localized states, if available. As was mentioned above,
there are two types of such states, namely CFR and DBH-type
levels. For the CFR states, one may, to a good approximation,
neglect the potential impurity scattering. To take properly into
account the Coulomb blockade effect on the impurity site, one has
to calculate the Green function in a self-consistent way, known as
"Hubbard I" approximation (see Eqs. (\ref{detB}), (\ref{occup}) in
the Appendix).

As was pointed out above, the origin of the energy gain is the
inter-impurity level splitting. The latter is derived in the
Appendix (Eqs. (\ref{deep1}) and (\ref{detmBa})). The result is
\begin{equation}\label{level1}
\delta E_{CFR\pm} = \pm \Delta  + \frac{9 K^2}{2} \frac{V^4}{[1 -
K V^2 P_{11}']^2}\frac{d P_{12}^2}{d\varepsilon},
\end{equation}
where the splitting of the levels of the isolated impurities is
\begin{equation}\label{level2}
\Delta({\bf R}) = \frac{K V^2 P_{12}({\bf R})}{1 - K V^2 P_{11}'}.
\end{equation}
The second term in Eq. (\ref{level1}) results from the repulsion
of the two-impurity levels from the band continuum. All the
functions $P_{ij}$ and their derivatives $P^\prime_{ij}$ are taken
at $\varepsilon = E^0_{CFR}$, i.e. at their positions for the
isolated impurities (\ref{deep1}).

Considering the contribution of the DBH states we may apply
the procedure outlined in the Appendix and to find the kinematic
exchange due to the DBH states lying above the top of the valence
band
\begin{equation}\label{DBH2a}
\Delta E_{DBH,ex} =  \frac{K V^2 P_{12}^2}{(E_{DBH}^0 - E_d -
V^2 P_{11}) P_{11}'}.
\end{equation}
Now all the functions in Eq. (\ref{DBH2a}) should be calculated at
the DBH level ($\varepsilon = E_{DBH}^0$). This part of the
kinematic exchange favors the ferromagnetic ordering, since
$P_{11}'(E_{DBH}^0) < 0$. If the DBH levels lie not too far from
the top of the valence band their contribution may be comparable
with that of the hole pockets and should be properly taken into
account when calculating the Curie temperature. At a sufficiently
high Mn content the splitting of the DBH level may result in the
formation of an impurity band and its merging with the {\em hh}
band. This case is discussed in the next section.

Thus one sees that the contribution, $\Delta E_{loc}$, of the
localized states to the kinematic exchange {\em is not universal}.
It depends on the type of conductivity and should be considered
separately for $p$- and $n$-type materials. In the $p$-type
samples one should take into account the empty shallow Mn-related
levels, which are present in $p$-(Ga,Mn)As and $p$-(Ga,Mn)P and
apparently absent in $p$-(Ga,Mn)N. In the $n$-type (Ga,Mn)N, the
localized states are due to the deep Mn CFR levels, and hopping
over these levels alone determines the exchange mechanism. The
double exchange mechanisms in these two cases are obviously
different.

To calculate the Curie temperature $T_C$ for the {\em{p}}- and
{\em{n}}-type materials one should extend the two-impurity
calculations for dilute materials with small Mn concentration $x$
and take into account the corresponding transformations in the
magnetic energy. This program is realized in the two next
sections.

\section{Magnetic order in {\em{p}}-type DMS}

\begin{figure}
\includegraphics[width=12cm]{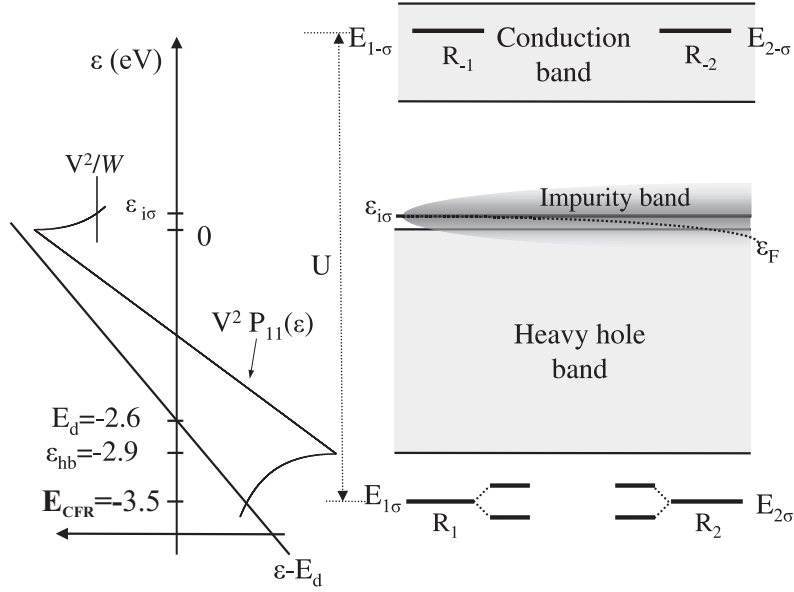}
\caption{Left panel: graphical solution of Eq. (\ref{deep}) for
the bonding CFR and Eq. (\ref{DBH0}) for the antibonding DBH
levels. Right panel:\ energy levels in (Ga,Mn)As. The CFR d-levels
$d^{5}/d^{4}$(denoted by R$_{1(2)}$) of each impurity, lie below
the \textit{hh } band. The DBH\ levels (energies $
\protect\varepsilon _{1\protect\sigma }$, $\protect\varepsilon
_{2\protect \sigma }$) are split from the \textit{hh} band and
form localized (acceptor) levels in the energy gap. The CFR levels
$d^{6}/d^{5}$ R$_{-1(-2)}$ lie high in the conduction band. The
impurity band is shown by the shaded area together with the
position of the Fermi level as a function of the width of the
impurity band (horizontal axis)} \label{f.1}
\end{figure}

Here we discuss the formation of ferromagnetic order in Mn-doped
III-V {\em{p}}-type semiconductors. (Ga,Mn)As together with
(Ga,Mn)P are the most celebrated among them. As mentioned above,
the CFR level of the fifth $d$ electron lies in these systems
below the heavy hole band. A DBH level is also formed above the
top of the valence band. At sufficiently high concentrations these
DBH levels start broadening into an impurity band, and may merge
with the valence band. The band structure of (Ga,Mn)As is shown
schematically in Fig. \ref{f.1}.

The empty localized levels lying above the Fermi energy. (While
making fitting of the theory to experimental data we use the Fermi
energy $\varepsilon_F$ instead of the chemical potential $\mu$,
since they nearly coincide under the condition of the experiment.)
appear due to the combined action of both the potential ($W$) and
resonance ($V$) scattering mechanisms. Formally they can be found
as zeros of the determinant $\mathsf{R}(\varepsilon)$ (see Eq.
(\ref{detB})). Neglecting small corrections due to resonance
scattering, the energy of an isolated DBH level,
$E_{DBH}^{0}\equiv \varepsilon _{i}$, corresponds to a zero of the
function
\begin{equation}
q(\varepsilon_i)=1-WP_{11}(\varepsilon_i)=0, \label{DBH0}
\end{equation}
like in the Koster-Slater one-impurity problem (see the graphic
solution on the left panel of Fig. \ref{f.1}). The calculations
below will be carried out assuming the model semi-elliptic density
of states for the {\em hh} band
$$\rho(\varepsilon) = \frac{2}{\pi w^2}\sqrt{w^2 - 4\left(
\varepsilon-\frac{w}{2} \right)^2},$$
where $w$ is the {\em hh} bandwidth and the energy $\varepsilon$
is counted from the top of the {\em hh} band. Then the position of
the acceptor DBH level can be found explicitly
$$
\varepsilon_i =E_{DBH}^{0}=W \left(1 - \frac{w}{4W}\right) ^{2}.
$$
Using a realistic value $w=2.9eV$ \cite{Mod} for the {\em hh}
bandwidth and taking the matrix element of the potential
scattering $W=1.02$ eV, one obtains the acceptor level position
$\varepsilon_i = 85meV$ for (Ga,Mn)As. The influence of resonance
scattering on the positions of the shallow DBH levels of the two
interacting impurities may be usually neglected (see Ref.
\onlinecite{Fleurov2} for a detailed description of the interplay
between resonance and potential impurity scattering).

At a sufficiently high impurity content, $x$, an impurity band is
formed in (Ga,Mn)As since the DBH levels are split due to the
indirect interaction within the impurity pairs. We neglect the
$pd$-hybridization $V$, keep only the contribution of the
potential scattering ($W$) and obtain
\begin{equation}\label{splitq}
\delta\varepsilon_\pm = - \frac{1}{W P_{11}~'} \textsf{q}_\pm,
\end{equation}
where $\textsf{q}_\pm$ are the two solutions of Eq. (\ref{DBH1}).

Then Eqs. (\ref{splitq}) and (\ref{DBH1}) result in
$$\Delta_{DBH}({\bf R}) \equiv \frac{1}{2}(\delta\varepsilon_+
-\delta\varepsilon_-) = -\frac{P_{12}({\bf R})}{P_{11}~'}$$
The functions $P$ and $P'$ are calculated at $\varepsilon =
\varepsilon_i$. The half-width of the impurity band can be roughly
estimated as $z \Delta_{DBH}(\overline{R})$ where $\overline{R}$
is the typical distance between the impurities, which depends on
the impurity content $x$. The $z$ value characterizes the number
of neighboring impurities participating in the interaction with
any given impurity. If the half-width $z\Delta_{DBH}(x)$ of the
impurity band exceeds the energy $\varepsilon_i$ (counted from the
top of the valence band), the impurity band merges with the
valence band and they both form a unified continuum of states. For
the above mentioned values of the model parameters this merging
occurs even at $x < 0.01$.

Eventually, the magnetic order is due to the exchange interaction
between the occupied CFR levels. These levels correspond to the
states $d^5/d^4$ of the Mn ions below the {\em hh} band, whereas
the empty $d^6/d^5$ CFR levels are pushed up to the conduction
band by the Anderson-Hubbard repulsion $U$ (Fig.~\ref{f.1}).

The left panel of this figure depicts a graphical solution of the
equation
\begin{equation}\label{sec2}
    {\sf R}(\varepsilon) = 0 ,
\end{equation}
with {\sf R}$(\varepsilon)$ defined in Eq. (\ref{detB}).
Neglecting the interaction between the impurities, $L_{12} = 0$
(and, hence $M_{12}(R)=0$), Eq. (\ref{sec2}) takes the form
\begin{equation}
E_{CFR}\equiv E_{i\sigma}=E_d +V^{2}L_{ii}(E_{i\sigma}),
\label{deep}
\end{equation}
and describes the formation of the deep impurity states of CFR
type out of the atomic d-levels $E_d$ below the {\em hh} valence
band. The values of the parameters used in the graphical solution
presented in Fig. \ref{f.1} will be discussed below.

For a finite distance between the impurities, the overlap of
impurity wave functions due to the nonzero $L_{12}(R)$ leads to a
level splitting as shown in the right panel of Fig.~\ref{f.1}.
Each of the two impurity levels is singly occupied due to the
Anderson-Hubbard on-site repulsion. The inter-impurity overlap
arises due to the $pd$-hybridization, which does not involve
spin-flips, hence it occurs only for the parallel impurity angular
moments (Hund rule). There is no level splitting for the
non-parallel impurity angular moments, unless one takes into
account indirect interaction via empty $d^6$ states.

FM ordering arises, provided the state with the parallel impurity
angular moments is energetically preferable in comparison to that
with the non-parallel ones. It is obvious that the splitting per
se cannot give an energy gain, when both states are occupied. One
should take into account all the changes in the energy spectrum,
namely, the reconstruction of the {\em{partially filled}} merged
impurity and {\em{hh}} band. The indirect interaction involving
empty states near the top of the {\em{hh}} band is in fact {\em {a
novel type of the double FM exchange}}, which resembles the well
known Zener exchange \cite{Zener} but differs from it in many
important aspects (see below).

The contribution favoring the FM order can be obtained from Eq.
(\ref{calc7}) with the addition of the part due to the impurity
band merged with the valence band,
\begin{equation}
\Delta E_{FM}=-\frac{V^{4}}{\pi }\int_{\varepsilon
_{F}(x)}^{\varepsilon _{i} + z \Delta_{DBH}(x) }d\varepsilon
\frac{\Gamma _{12}(\varepsilon )P_{12}(\varepsilon )}{[\varepsilon
-E_{d}-V^{2}P_{11}({\varepsilon })]^{2}+\frac{V^{4}}{4}\Gamma
_{11}^{2}\left( \varepsilon \right) },  \label{Ehol}
\end{equation}
(the reference energy in this equation is taken at the top of the {\em hh}
band). In the FM-aligned spin configuration only the {\em
majority} spin subband contributes to $\Delta E_{FM}$, therefore
the spin index is omitted for the sake of brevity.

We use in our numerical estimates the following approximate
relations,
\begin{eqnarray}
P_{11}(\varepsilon ) &=& P_{22}(\varepsilon )=\int d\omega
\frac{\rho (\omega )}{\varepsilon -\omega },\;\;P_{12}(\varepsilon
)=\int d\omega \frac{\sin [k\left( \omega \right) R]}{k\left(
\omega \right) R}\frac{\rho (\omega )}{\varepsilon -\omega },
\nonumber \\
\Gamma _{12}(\varepsilon ) &\approx &2\pi \rho (\varepsilon )\sin
[k\left( \varepsilon \right) R]/k\left( \varepsilon \right) R,\ \
\ \Gamma _{11}(\varepsilon )\approx \pi \rho (\varepsilon )=\Gamma
_{22}(\varepsilon ).  \label{selfi}
\end{eqnarray}
(here $R$ is the inter-impurity distance).  The value of the
wave-vector $k$ is found from the equation $\varepsilon
=\varepsilon _{hh}(k)$, where $\varepsilon _{hh}(k)$ is the
\textit{hh} energy dispersion. In our model calculations the
impurity band merges with the top of the valence band of (Ga,Mn)As
at the concentration $x_{crit}=0.0065$ with the renormalized
3d-wave-function enveloping 5 to 6 unit cells. Earlier transport
measurements for low $T_{C}$ samples indicate merging even at
$x=0.002$. All this justifies the approximation adopted for the
hole state in Eq. (\ref{Ehol}). This equation is our working
formula, from which we obtain $T_C$.

The mechanism outlined above competes with the antiferromagnetic
(AFM) Anderson-type superexchange \cite{Alex} involving the empty
$d^6$ states. The energy gain in the latter case is estimated as
\begin{equation}
\Delta E_{AFM}\approx - V^{4}L_{12}^{2}/U\;.  \label{iae}
\end{equation}
It should be compared with the above Zener-type mechanisms. We
assume that
$$P_{ij}\sim w^{-1},~~~\Gamma _{ij}\sim \varepsilon
_{F}^{1/2}w^{-3/2},~~~ \varepsilon _{F}-E_{d}-V^{2}P_{11}({
\varepsilon }_{F})=4\alpha V^{2}/w\;,$$
with $\alpha <1$ (see left panel of Fig. \ref{f.1}). Then one
finds from Eqs. (\ref{Ehol}) and (\ref{iae}) that $|\Delta
E_{AFM}|\;\sim V^4 a^2 \exp \left( -2 \kappa _{b}R_{12}\right)
/(Uw^2R_{12}^2)$, with $\kappa_b =\sqrt{ 2m\left( \varepsilon
_{b_h}-E_{d\sigma }\right)}/\hbar $, $|\Delta E_{FM}|\;\sim
2\varepsilon_F\left( \varepsilon _{F}/w\right) ^{1/2}/\left[
\left( 4\alpha \right) ^{2}+\varepsilon _{F}/\left( 4w\right)
\right] $ and FM coupling is realized provided $|\Delta
E_{FM}|>|\Delta E_{AFM}|$, which normally takes place since the
parameter $U$ is large. Numerical estimates are given below.

Now the difference between the double exchange mechanism in DMS
and the Zener double exchange in transition metal oxides becomes
clear. The conventional Zener double exchange mechanism was
proposed\cite{Zener} for (La, A$^{2+}$)MnO$_{3}$. In this case Mn
ions are in different valence states (Mn$^{3+}$ and Mn$^{4+}$). In
our case it would have meant that one of the two levels
$E_{CFR\uparrow}$ were empty. It is seen from Fig. \ref{f.1} that
in (Ga,Mn)As both these states are occupied, and the Zener
mechanism in its original form is not applicable. Actually, the FM
order in the {\em{p}}-type DMS arises due to the kinematic
exchange between the two Mn(d$^5$) ions via the empty states near
the top of the valence band.

In order to calculate the exchange energy Eq. (\ref{Ehol}), the
dependence of the Fermi level on the Mn content $\varepsilon_F(x)$
should be known. It is found from the equation
\begin{equation}\label{con}
x_{s}=2\int\limits_{\varepsilon _{F}(x)}^{\delta w( x)}\rho \left(
\varepsilon \right) d\varepsilon,
\end{equation}
$\delta w( x)= \varepsilon_i+z\Delta_{DBH}(x)$, and the {\em hh}
density of states is approximated as
$$
\rho \left( \varepsilon \right) =8/ \left[ \pi \left(w+ \delta w(
x) \right) ^{2}\right] \sqrt{\left[ \delta w( x) -\varepsilon
\right] \left( \varepsilon +w\right)} \theta \left[ \delta w( x)
-\varepsilon \right] \theta \left( \varepsilon +w\right).
$$
($z$ is the coordination number in the impurity band). The per
site hole concentration $x_s$ is proportional to the hole density
$p_h$; $x_s = a^3 p_h/8$ in the zinc-blende structure. Equations
(\ref{Ehol}) and (\ref{con}) allow  one to determine the pair
exchange energy as a function of Mn concentration $x$ and connect
it with the experimental data for the given hole concentrations
$p_h(x)$ which are taken from the measurements. \cite{Gall}

The Curie temperature $T_C$ was calculated in the molecular field
approximation. According to this approach the spin Hamiltonian reads
\begin{equation}\label{tc}
H_{MF}=\frac{1}{2}\sum\limits_{i}\Delta E_{FM}({\bf R}_{ij}){\bf
J}_{i}\cdot \left\langle {\bf J} \right\rangle \;,
\end{equation}
where summation runs over all positions of the Mn impurities with
an angular moment operator ${\bf J}_i$. Here the factor 1/2
accounts for the fact that the energy gain of the FM vs AFM
orientation of two coupled spins is $\Delta E_{FM}({\bf R}_{ij})$,
contrary to $2J_{Heis}$ in the Heisenberg model. Then the Curie
temperature can be readily found,
\begin{equation}\label{Curie}
T_{C}=- \frac{\Delta E_{FM}(x)}{k_B} \frac{\overline{z}
J(J+1)}{6}.
\end{equation}
where $\overline{z}$ is, similarly to $z$, a measure of the
neighboring atoms participating in the exchange interaction. It is
certainly close to $z$, although it does not necessarily coincide
with it. It is worth mentioning here that unlike the case of
magnetically doped metals \cite{Caroli} we cannot represent the
effective inter-impurity coupling in the asymptotic form
$$
\Delta E_{FM}(R_{ij}) \sim cos (2k_FR_{ij}+\phi)/(k_FR_{ij})^3
$$
because typically $k_F ^{-1}\sim FR_{ij}$ in doped semiconductors.

\begin{figure}[th]
\includegraphics[width=14cm]{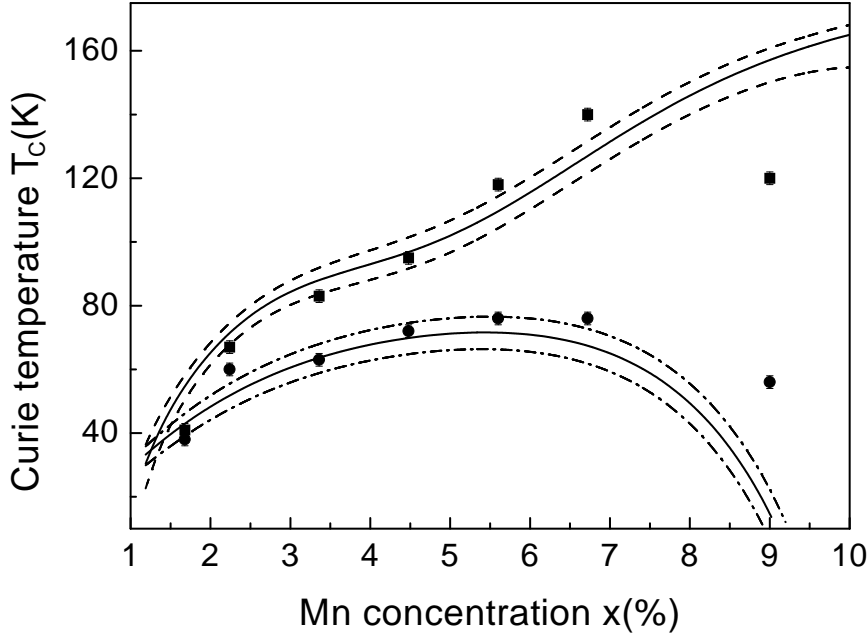}
\caption{Calculated dependence of $T_C$ on the manganese
concentration $x$ in (Ga,Mn)As based on the experimental data for
the hole density\cite{Gall} $p_{h}\left( x\right) $. Solid squares
(circles) stand for experimental $T_{C}\left( x\right)$ of
annealed (as-grown) samples. Broken lines take into account the
error bars of the hole density.} \label{f.2}
\end{figure}

\begin{figure}[th]
\includegraphics[width=14cm]{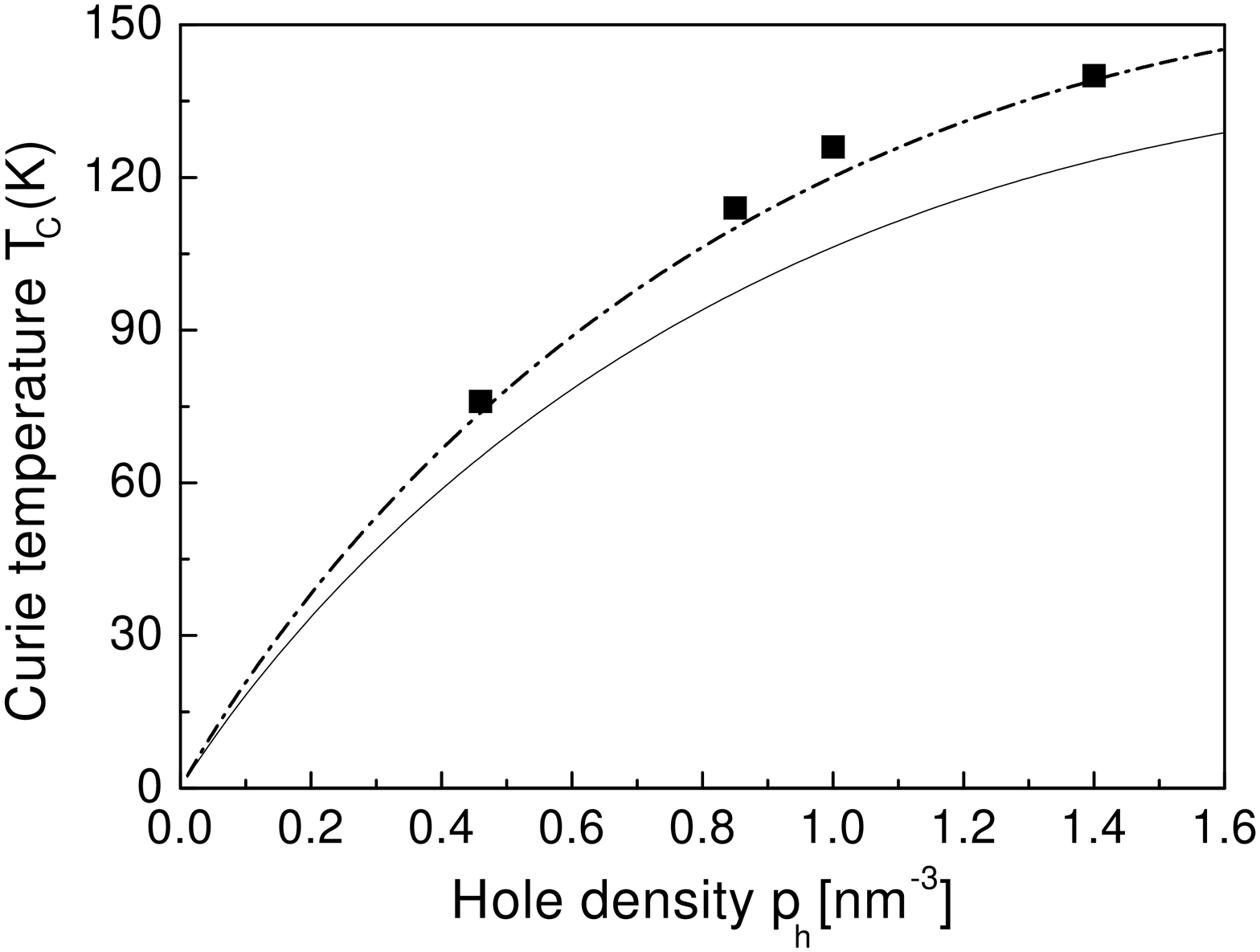}
\caption{The dependence of the Curie temperature on the hole
density $p$ in (Ga,Mn)As. Solid and dash-dotted lines are the
theoretical curves for the two values of the hybridization
parameter ($V=1.22$ and 1.24 eV, respectively). Experimental
points (filled squares) are taken from Ref. \onlinecite{Gall}.}
\label{f.3}
\end{figure}

In (Ga,Mn)As the total angular moment of a complex
Mn($3d^{5}\overline{p}$) is unity \cite{Linnarson}: $J=1$, since
it is formed by the moment $j=3/2$ of the loosely bound hole
AFM-coupled to the Mn center with the spin $S=5/2$. Then the
numerical factor in Eq. (\ref{Curie}) is close to unity. The
results of our calculations are presented at Figs. \ref{f.2} and
\ref{f.3}.

In order to find the dependence $T_C(x)$, the calculated value of
the Fermi level via Eq. (\ref{con}), based on the experimental
data of Ref. \onlinecite{Gall}, is used in Eqs. (\ref{Curie}) and
(\ref{Ehol}). A polynomial fit is used for the hole density
$p_h(x)$ in Eq. (\ref{Ehol}). The values of the model parameters
characterizing the impurity $d$-state are $U\approx 4.5$ eV,
$V=1.22$eV, while the {\it hh} mass  $m=0.51\cdot m_0$ and {\it
hh} bandwidth $w=2.9 eV$ were taken from Ref. \onlinecite{Mod}.
The hybridization strength $V$ was obtained from Eq. (\ref{deep})
with $E_{CFR}=-3.47$eV ($E_{CFR}^{\exp }=-3.4$eV according to Ref.
\onlinecite{Asklund}). The value of $ \varepsilon _{i}=85$ meV was
chosen for the energy of the shallow acceptor level (not far from
the experimental value\cite{Zunger} $\varepsilon _{i}^{\exp }=110$
meV.) The value of $\bar{z}=2.5$ was taken for the coordination
number in as-grown samples (Fig.2, lower curve). At these values
for the model parameters the ratio $ |\Delta E_{FM}|/|\Delta
E_{AFM}|\sim 2$ justifies the dominance of the FM coupling in
(Ga,Mn)As.

The non-monotonous dependence $T_C(x)$ is due to a non-equilibrium
character of the sample preparation. Apparently, the ratio between
the concentration of Mn impurities and the actual hole
concentration depends on the doping method and the thermal
treatment. In particular, the annealing of the sample results in a
reduction of the donor-like Mn-related interstitial defects in
favor of acceptor-like substitution impurities.\cite{Gall,Yu} To
describe the annealing effect we changed the value of $\bar z$
from 2.5 to 4 in Eqs. (\ref{Ehol}) and (\ref{Curie}). The results
of numerical fitting are shown in the upper curve of Fig.
\ref{f.2}.

\begin{figure}[th]
\includegraphics[width=14cm]{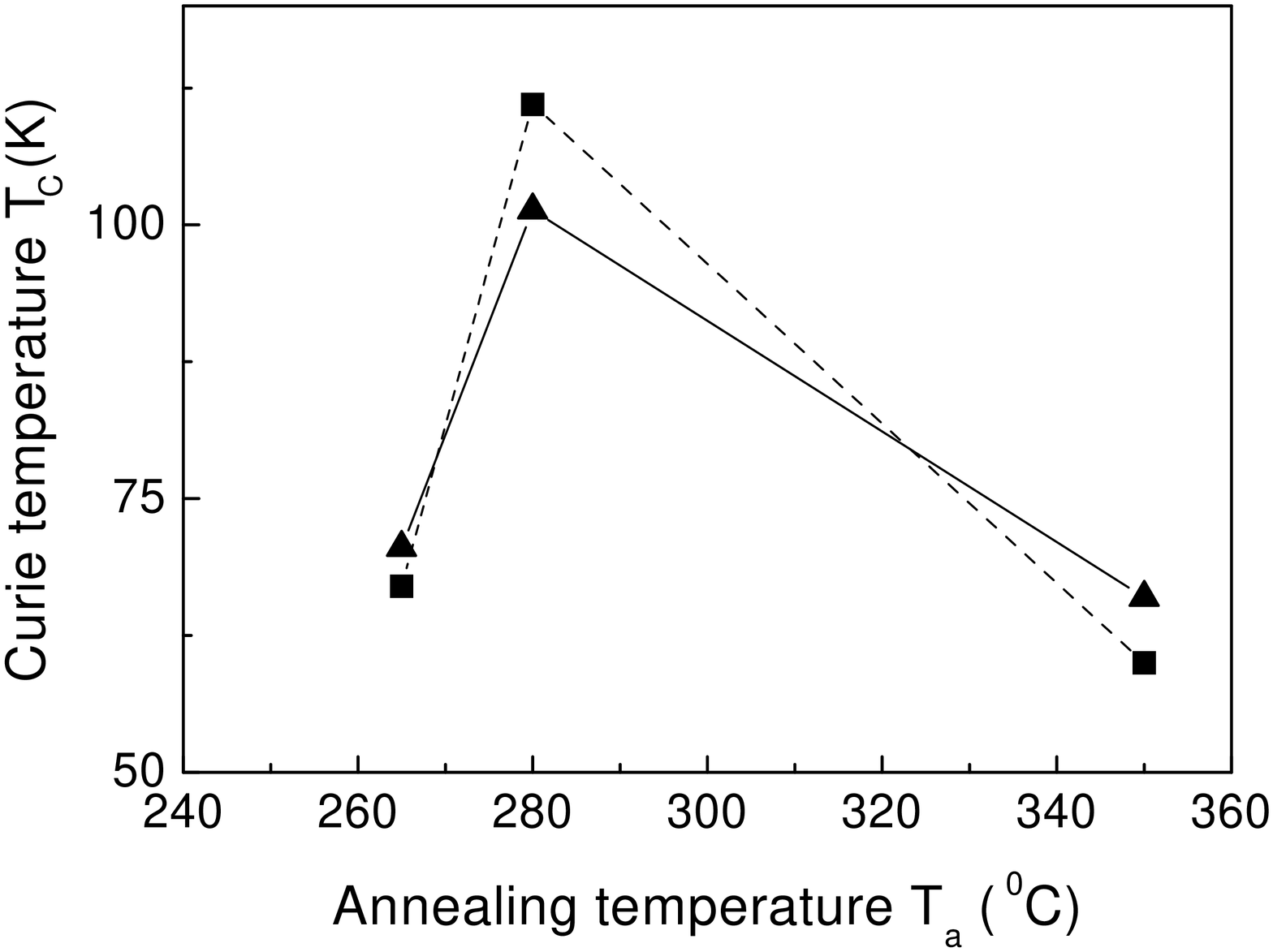}
\caption{Calculated $T_{C}$\ (triangles connected by a solid line)
{\em vs} the annealing temperature $T_a$ in (Ga,Mn)As. The data
for the hole density and the experimental values of $T_C(T_a)$
(closed squares) are taken from Ref. \onlinecite{Yu}. The
parameters used in the calculations are: $V$ $=1.22eV,$
$\bar{z}=4.0,E_{CFR}=-2.6eV$.} \label{f.4}
\end{figure}

Recent detailed measurements of the hole concentrations in a
series of both as-grown and annealed  Ga$_{1-x}$Mn$_x$As samples,
\cite{Gall} allowed us to compare the theoretical plot $T_C(p)$
with the experimental data. These results are presented in Fig.
\ref{f.3}. Our fitting procedure uses the same equations
(\ref{Curie}) and (\ref{Ehol}), and the same values for the model
parameters $E_{CFR}$, $U$, $\varepsilon_i$ as in the above
estimate. The Mn concentration point $x=0.067$ is used as a
reference point, and the coordination number $\bar{z}=4$ is
chosen. Two theoretical curves correspond to two values of the
hybridization parameter $V=1.22$ and 1.24 eV (solid and
dash-dotted curves, respectively). One may conclude from these two
fittings that the theory is not very sensitive to the choice of
the model parameters. To check this statement, we made one more
fitting of the experimental data obtained only for annealed
samples. (see Fig. \ref{f.4}). These data are taken from Ref.
\onlinecite{Yu}. We see that the calculations with the same set of
model parameters give satisfactory quantitative agreement with the
experiment for these samples as well.

Thus, our theory provides a satisfactory quantitative description
of the behavior of the Curie temperature $T_C$ as a function of
the Mn content $x$ and the hole concentration $p_h$ in
$p$-(Ga,Mn)As. This description is applicable also to
$p$-(Ga,Mn)P. The band structure of (Ga,Mn)P can be also
schematically represented by Fig.1, however, the DBH level lies
deeper in the energy gap than in $p$-(Ga,Mn)As: its position is
estimated as $\approx 0.4$ eV above the top of the valence band
(see, e.g. Refs. \onlinecite{Kikoin,Zunger,graf,Mod}). The
impurity band appears to be narrower than in (Ga,Mn)As. It does
not merge with the {\em hh band} at any reasonable Mn
concentration and its contribution to the indirect exchange can be
neglected. We tested our theory by fitting the experimental
value\cite{hebard} of $T_C=270$ K reported for a {\em p}-type
(Ga,Mn)P at $x=0.06$. The impurity level $\varepsilon_i = 400$ meV
arises at $W=1.4$ eV. We have found that the value of $T_C=270$K
is reproduced by means of Eq. (\ref{Curie}) with the following
fitting parameters: $w=2.6$ eV, $V=1.31 $ eV, $x=0.05$, $\bar z
=4$, $E_d=-1.3$ eV (then $E_{CFR}=-3.28$ eV). The chosen value of
$p_h=1\cdot 10^{20}$ cm$^{-3}$ is the hole density in a C-doped
GaP before Mn implantation.\cite{hebard} So the values of our
model parameters are close to those for (Ga,Mn)P in the
experiments of Ref. \onlinecite{hebard}. Besides, these parameters
are in a good correlation with the chemical trends in the systems
(Ga,Mn)As, (Ga,Mn)P (the latter has a smaller lattice spacing and
larger effective mass $m_{hh}^*$ than the former).

As for {\em{p}}-(Ga,Mn)N, the theory should be somewhat modified
in order to describe this material. In this case Mn remains a
neutral substitution isoelectronic defect in the configuration
Mn$^{3+}$($d^4$), since the Mn$^{2+/3+}$ transition energy falls
deep into the wide energy gap in accordance with the experimental
\cite{graf} and theoretical \cite{Sato} data. If the hole
concentration, $p_h$, exceeds the Mn content, $x$, then we return
to the situation discussed in this section, but with the localized
moment $J=2$ characteristic of the Mn$^{3+}$($e^2t^2$)
configuration. In the opposite case, the valence band is full, and
the Fermi energy lies within the deep impurity band formed by the
Mn$^{2+/3+}$ levels. This situation is discussed in the next
section.

Unfortunately, there are not enough experimental data on $T_C(x)$
available to make a detailed comparison with the theoretical
predictions both in {\em{p}}-(Ga,Mn)P and {\em{p}}-(Ga,Mn)N
materials. Moreover, no $p$-type conductivity was observed in
(Ga,Mn)N even in the cases when the pristine GaN crystals were of
a {\em p}-type. Due to the lack of experimental information, we
have confined ourselves with a quantitative description of the
{\em p }-type (Ga,Mn)As and turn now to the case of {\em
n}-(Ga,Mn)N, where the double exchange mechanism, as described
above, should be revisited.

\section{Magnetic order in {\em{n}}-type DMS}

In accordance with the general predictions of the theory,
\cite{Kikoin,Zunger} Mn is a deep acceptor in GaN, and each Mn
impurity creates an {\em empty} $t_2$ level close to the middle of
the energy gap of GaN. Therefore, the Fermi energy should be
pinned to these levels unless other (shallow) acceptors create
enough free holes near the top of the valence band. This statement
is confirmed by extended cluster "quasi band"
calculations.\cite{Sato}  If a Mn-doped sample contains also
shallow donors in a noticeable concentration, then the deep
Mn-related impurity band is partially filled by electrons, and one
arrives at the problem of magnetic order in {\em n}-type (Ga,Mn)N.

In this case the $d^5/d^4$ CFR energy levels forming the impurity
band are partially filled, so one encounters the mixed valence
Mn$^{3+}$/Mn$^{2+}$ situation similarly to the original Zener
model.\cite{Zener} Hopping in the impurity band is possible due to
an overlap of the tails of the impurity wave functions. These
tails are formed by the superpositions of the Bloch electron wave
functions from the $hh$ band, so the latter in fact plays the role
which is similar to that of the oxygen p-orbitals in
La(Mn,Sr)O$_3$ considered by Zener.

\begin{figure}[h]
\includegraphics[width=14cm]{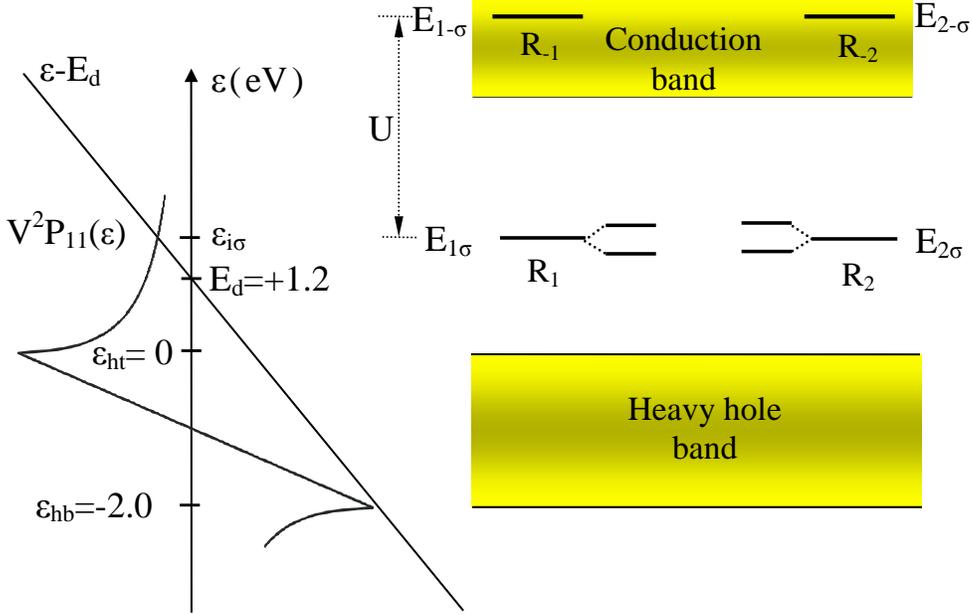}
\caption{Left panel: graphical solution of Eq. (\ref{deep}) for
the antibonding CFR level in (Ga,Mn)N. Right panel:\ energy levels
in Ga(Mn)N. The CFR d-levels $d^{5}/d^{4}$ (denoted by R$_{1(2)}$)
of each impurity, lie within the energy gap. The CFR levels
$d^{6}/d^{5}$ R$_{-1(-2)}$ lie high in the conduction band.}
\label{f.6}
\end{figure}

The specific feature of the {\em n}-type (Ga,Mn)N DMS's is that
the $d^5$/$d^4$ CFR impurity levels, $E_{CFR}$, lie deep within
the forbidden energy gap. At a high enough Mn content they may
form an impurity band, which still remains well separated from the
{\em hh} band. This impurity band may be partially occupied by
electrons and be responsible for the possible magnetization of the
DMS. To describe formation of a deep impurity band within the
framework of our model, we solve Eq. (\ref{deep}) for a Mn-related
CFR level and then substitute this solution into the secular
equation (\ref{sec2}). These calculations are performed by using
the same semi-elliptic density of valence states
$\rho(\varepsilon)$ as in Section III. The bare level $E_d$ in
this case is above the top of the $hh$ band, and the $pd$
hybridization only slightly renormalizes its position in the
energy gap. The graphical solution of Eq. (\ref{deep}) is
presented in the left panel of Fig. \ref{f.6}.

We assumed that the hybridization parameter is nearly the same as
in (Ga,Mn)As, $V=1.2$ eV. The right panel of Fig. \ref{f.6}
illustrates the formation of an impurity dimer via an overlap of
the tails of the CFR wave functions in accordance with Eq.
(\ref{sec2}). Then the electron hopping between the correlated
states $R_1$ and $R_2$ initiates formation of an impurity band and
the Zener type indirect exchange interaction arises, provided the
impurity band is partially occupied. In principle, the double
exchange via empty states in conduction band should be taken into
account. However, this contribution is small in zinc blende host
crystals, because the hybridization $V_{ds}$ between the impurity
$d$-states and the $s$-states of the electrons near the bottom of
conduction band is weak due to the symmetry selection
rules.\cite{Kikoin,Zunger} The empty $(d^6/d^5)$ CFR levels are
pushed up high to the conduction band by the Anderson-Hubbard
repulsion $U$. Like in the previous case these levels can be a
source of AFM ordering in the impurity band, but we do not
consider this relatively weak mechanism here. Another contribution
to indirect exchange in the case of $n$-(Ga,Mn)N is the
fourth-order in $|V_{dp}|^2|V_{ds}|^2$ Bloembergen-Rowland
indirect exchange. \cite{Barz} However, this term involves the
hybridization with both valence and conduction electron states, so
it is small due to the above symmetry ban.

In order to calculate the double exchange interaction in such an
impurity band, we consider first a pair of impurities at a
distance $R$ from each other. To be definite we assume that the
chemical potential $\mu$ lies below the level $E_{CFR}$ of an
isolated impurity. According to Eq. (\ref{level2}) the interaction
between the impurities results in splitting of the impurity levels
with half-width $\Delta(R)$, where we neglect its possible
dependence on the direction of the vector {\bf R}.

It is emphasized here again that the $pd$ hybridization does not
involve spin flips. It means that the electron spin does not
change its projection in the course of the whole process leading
to the indirect exchange, during which the electron is transferred
from one impurity to the other and back via band states. This
process is possible only if the spins ${\bf S}_i$ of the two
impurities are parallel. Otherwise the Hund rule will require an
additional energy when the electron tries to hop onto the impurity
atom with a "wrong" spin orientation.

One may conclude that the indirect exchange interaction within a
pair of impurities appears if their angular moments are parallel
(FM ordering), and if the distance between the impurities is not
too large, i.e.
$$R < R_{ex}(\mu)$$
where $R_{ex}(\mu)$ is the solution of the equation
\begin{equation}\label{critrad}
\Delta(R_{ex})= E^0_{CFR} - \mu.
\end{equation}
Then one of the two impurity levels sinks below the chemical
potential $\mu$ and is occupied by an electron resulting in the
energy decrease $\Delta(R)$, whereas the second level remains
empty and does not contribute to the energy balance. If the
angular moments of the impurities are not parallel, Hund rule
forbids the indirect exchange and there is no energy decrease.
Therefore, we conclude that the exchange energy in such a pair
equals just $- \Delta(R)$. It is zero if at least one of the above
two conditions is not met.

The pair exchange interaction can be estimated as
\begin{equation}\label{avdelta}
\overline{\Delta} = \int_0^{R_{ex}(\mu)} \Delta(R) g(R) dR
\end{equation}
where $g(R)$ is the impurity pair distribution function. For
example, we may assume that the random distribution of the
impurities obeys the Poisson law
\begin{equation}\label{distr}
g(R) = \frac{3 R^2}{\overline{R}~^3} \exp \left \{-
\frac{R^3}{\overline{R}~^3} \right\}
\end{equation}
where $\overline{R}$ is the average distance between the
impurities. The latter is connected with the impurity content $x$
by $\frac{4\pi}{3} \overline{R}~^3 x = d_0^3$, where $d_0$ is the
minimal possible distance between two Mn ions. This assumption may
not work too well for relatively high impurity content, since it
overestimates the contribution of closely lying impurities. In
fact the impurities cannot approach each other to distances
smaller than the distance between two neighboring Ga atoms,
$a\sqrt{2}/2$, $a$ lattice constant. In such a case corrections
should be introduced for these small distances.

In order to complete our calculations, we need also an equation
connecting the position of the chemical potential with the
electron concentration $n$ in the impurity band. This
concentration coincides with the concentration of impurities in
the $d^5$ state, hence
\begin{equation}\label{conc1}
n = \int^\mu \rho(\varepsilon) n_{d^5}(\varepsilon - \mu)
d\varepsilon
\end{equation}
where $\rho(\varepsilon)$ is the density of states in the impurity
band. The distribution $n_{d^5}(\varepsilon - \mu)$ is calculated
in the Appendix, Eq. (\ref{occup}).

Since, in unfilled  impurity band the chemical potential and the
Fermi level $\varepsilon_F$ practically coincide, we use below the
latter term as in the case of $p$-type samples instead of $\mu$.
At low temperatures $n_{d^5}(\varepsilon - \varepsilon_F)$ is 1
for $\varepsilon < \varepsilon_F$ and 0 for $\varepsilon >
\varepsilon_F$. This allows us to circumvent the calculation of
the density of states $\rho(\varepsilon)$ and present the
condition (\ref{conc1}) in the alternative form
\begin{equation}\label{conc2}
n = \int_0^{R_{ex}(\varepsilon_F)} g(R) dR,
\end{equation}
which is more convenient for our calculations. If an experiment
provide us with dependencies of the Curie temperature and the
electron concentration on the Mn content, then Eqs.
(\ref{critrad}) and (\ref{conc2}) will allow us to connect the
Fermi energy with the electron density in the impurity band
and after that to use Eq. (\ref{avdelta}) in order to
calculate the average pair exchange interaction
$\overline{\Delta}(x)$ as a function of the impurity content $x$.

We may now use the theory of dilute ferromagnetic alloys developed
in Ref. \onlinecite{ks72} (see also the review
Ref.~\onlinecite{ks78}) and arrive at the final result of this
procedure, i.e. the Curie temperature. Unlike the case of
(Ga,Mn)As, the magnetic energy in impurity band is determined by
the spin $S$ of Mn $3d$ shell, which equals 5/2 and 2 for $d^5$
and $d^4$ configurations, respectively. In our crude approximation
it is sufficient only to know that the proportionality coefficient
between $T_C$ and $\bar \Delta$ is of order of unity:
$$T_C(x)= \frac{S(S+1)}{6 k_B} \overline{\Delta}(x) \approx \bar
\Delta(x)/k_B.$$

At low electron concentrations, when $R_{ex}(\varepsilon_F) >
d_0$, only a minor part of the impurities is coupled by the
indirect exchange interaction. Using the above averaging procedure
for calculating $T_C$ does not hold any more and a percolation
type of approach should be applied. The resulting Curie
temperature may become low and will decrease exponentially with
the concentration.\cite{kss73}

The behavior of the Curie temperature for higher electron
concentrations, when the Fermi energy lies above the energy
$E^0_{CFR}$, can be found in a similar fashion. We just have to
switch to the hole representation and use the distribution
function $3n_{d^4}(\varepsilon - \varepsilon_F)$ instead of
$n_{d^5}( \varepsilon - \varepsilon_F)$. (Dealing with holes we
have now to account for the three-fold degeneracy of $t_2$
states.) The function $3n_{d^4}$ is 1 if $\varepsilon >
\varepsilon_F$ and 0 if $\varepsilon < \varepsilon_F$. As a
result, we will obtain a mirror symmetric dependence of $T_C$ on
$x$ in this concentration range.

To get an idea of the behavior of the average kinematic exchange
$\overline{\Delta}$ and, hence of the Curie temperature as a
function of the Mn content and the position of the Fermi energy,
i.e., electron concentration in the impurity band, we can make
simple estimates. Consider the case when $\varepsilon_F <
E_{CFR}$, then $K=1$ and according to Eq. (\ref{detmBa}) the
levels of a pair of impurities separated by the distance $R$ are
split by
\begin{equation}\label{splitr}
\delta\varepsilon_\pm = \pm \Delta(R) =\pm \frac{V^2 P_{12}(R)}{1
- V^2 P'_{11}}
\end{equation}
where the off-diagonal lattice sum can be estimated as
\begin{equation}\label{lattice}
P_{12}(R) = \frac{1}{2\pi} \frac{m d_0^3}{\hbar^2 R} e^{-\kappa
R}.
\end{equation}
Here $\kappa$ is the localization parameter. (Although we may use
the estimate $\kappa \approx \hbar\sqrt{2 m E_{CFR}}$, we should
not forget that it may be rather crude for really deep levels.)
Then
$$\Delta(R)\approx \Delta_0 \frac{d_0}{R} e^{-\kappa R}\;, $$
where the value of $\Delta_0$ depends on the parameters of the
system but is generally of the order of 1eV.

\begin{figure}[th]
\includegraphics[width=15cm]{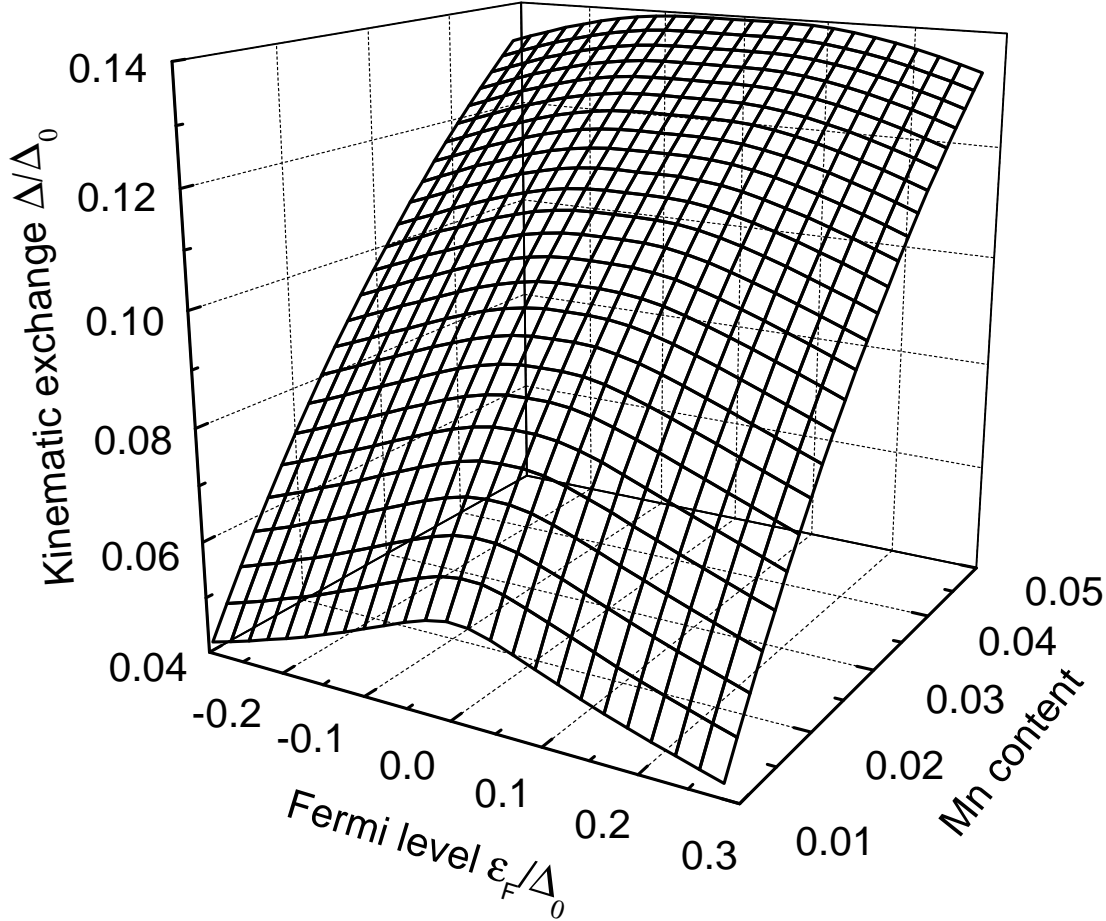}
\caption{Dependence of the kinematic exchange, measured in units
of $\Delta_0$ on the Mn content ($x$), and on the position of the
Fermi energy relative to the level $E_{CFR}$ of an isolated Mn
impurity. We took $\kappa^{-1} = d_0$.} \label{f.7}
\end{figure}

The Poisson pair distribution truncated at small distances $R<d_0$
is
\begin{equation}\label{distr1}
g(R) = \left\{ \begin{array}{cc} \displaystyle \frac{3
R^2}{\overline{R}~^3} \exp \left \{ \frac{d_0^3 -
R^3}{\overline{R}~^3} \right\},&~~~ \mbox{for} ~~~ R > d_0
\\
0,&~~~ \mbox{for} ~~~ R < d_0
\end{array}.
\right.
\end{equation}
This distribution accounts for the fact that the Mn impurities
cannot come closer than at the distance $d_0$ from each other. For
a deep enough level we may assume $\kappa\approx 1/d_0$ and
calculate the dependence of the average kinematic exchange
$\overline{\Delta}$ on the Mn content and position of the Fermi
energy with respect to the isolated impurity level $E_{CFR}$ (see
Fig. \ref{f.7}). The two latter quantities are measured in units
$\Delta_0$ meaning that the kinematic exchange and, hence, the
Curie temperature is maximal for any given value of $x$ when the
impurity band is half filled, i.e., $\varepsilon_F=E^0_{CFR}$. At
$x = 0.05$ the kinematic exchange $\overline{\Delta}$ may be about
0.14eV (if we assume that $\Delta_0 =1$eV). The angular moment
$J=2$ for the $d^4$ state of the Mn impurity in GaN, hence $T_C=
\overline{\Delta}/k_B$, and we find that the value of the Curie
temperature varies from a few hundred K for $x=0.01$ and may
exceed 1000K at $x=0.05$.

\begin{figure}[th]
\includegraphics[width=15cm]{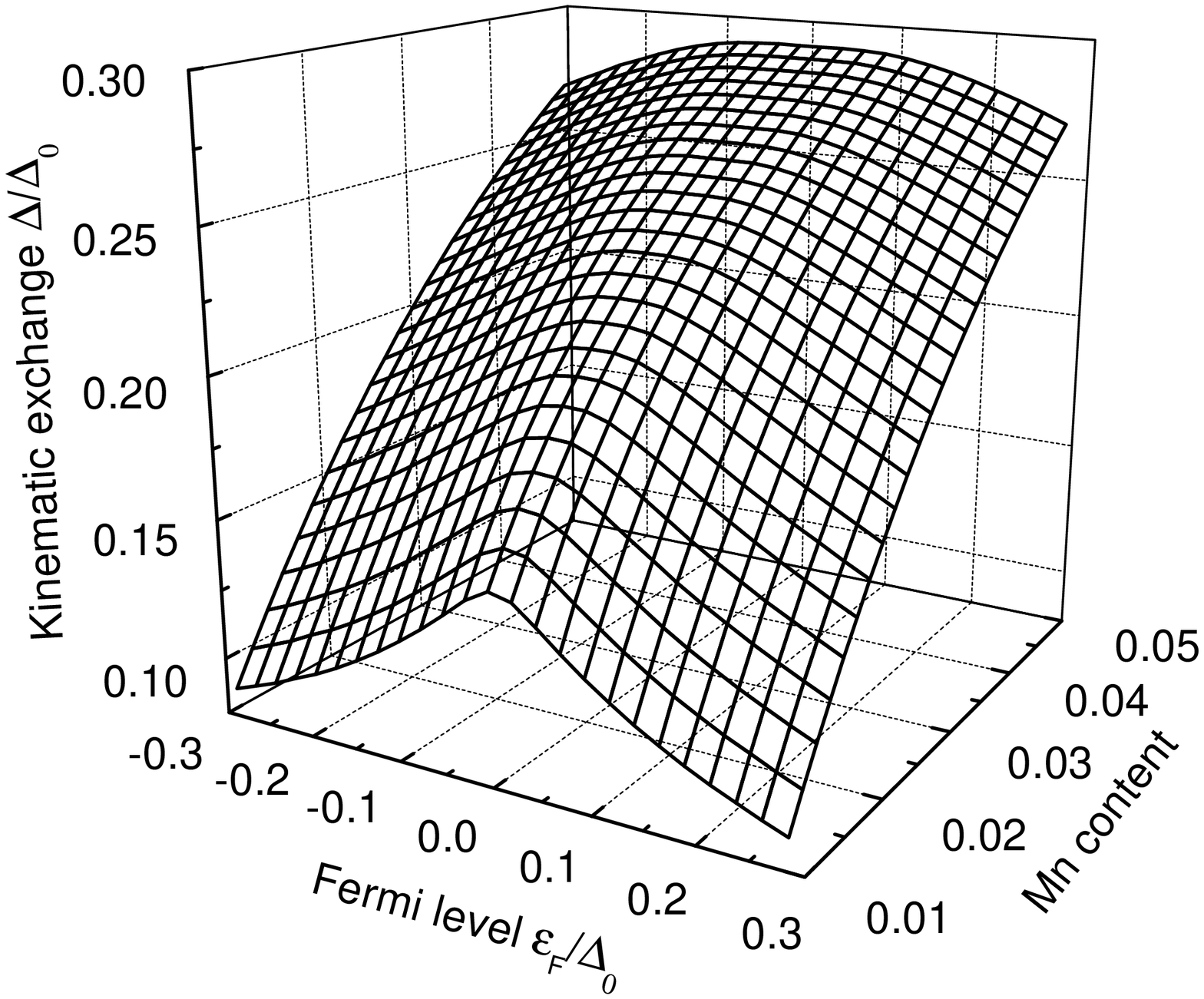}
\caption{The same as Fig. 6 but now for $\kappa^{-1} = 2d_0$.}
\label{f.8}
\end{figure}

It is worth mentioning that by choosing the localization radius so
small, $\kappa^{-1} \approx d_0$, we have actually found a lower
bound for $\overline{\Delta}$. At larger values of the
localization radius, we obtain a similar dependence of the
exchange energy $\overline{\Delta}$ on the Mn content and the
Fermi energy position, however the absolute value of
$\overline{\Delta}$ may become essentially larger, e.g. $\sim$0.3
for $\kappa^{-1} \approx 2d_0$ in the maximum (see Fig. \ref{f.8})
meaning that the Curie temperature may become as high as several
thousands K.


We should emphasize here again that the averaging procedure used
to obtain the results in Figs. \ref{f.6} and \ref{f.7} works well
only when the Fermi energy is close to the impurity level
$E^0_{CFR}$ (zero on the Fermi energy axis). Far from this region
one should apply the percolation approach proposed in Ref.
\onlinecite{kss73}. Then the effective kinematic exchange in this
range may become smaller than shown in the figures, however the
behavior does not change qualitatively, and certainly remains
correct also quantitatively for $\varepsilon_F$ close to
$E^0_{CFR}$.

To conclude this section, we have found that the optimum
$n$-doping level for FM alignment of Mn spins corresponds
approximately to half-filled impurity band, and one may well
expect really high temperature magnetism in this case. The Curie
temperature of $T_C =940$K, reported in Ref. \cite{sonoda}, lies
well within the range of values shown in Figs. \ref{f.6} and
\ref{f.7}. Unfortunately the lack of information on the impurity
concentration and position of the Fermi level does not allow us to
make a more detail comparison with the experiment.

\section{Concluding remarks}

According to the general theory of magnetic interactions, the
kinematic exchange between localized impurity spins embedded in a
semiconductor host with covalent chemical bonding\cite{coval} is a
generic property of the two-impurity cluster. We constructed such
a cluster which was based on an ab initio knowledge about the
electronic structure of isolated Mn impurity in a III-V
semiconductor. After some simplifications this approach led us to
a microscopic Hamiltonian similar to a two-center Anderson
model.\cite{Alex,Caroli} That model was developed for metallic
hosts, whereas we consider a semiconductor host. The model allows
for analytical solutions, and the final equation (\ref{Curie}) for
$T_C$ with the effective coupling constant (\ref{Ehol}) should be
compared with the phenomenological mean-field equation
\cite{McDo,Domcf}
\begin{equation}\label{Oh}
T_C=\frac{4x S(S + 1)J_{pd}^2}{3a}\frac{\chi_h}{(g\mu_B)^2}
\end{equation}
where $a$ is the lattice constant, $J_{pd}$ is the
phenomenological $pd$-exchange constant, $\chi_h$ is the magnetic
susceptibility of holes in the valence band. In principle, Eq.
(\ref{Curie}) and Eq. (\ref{Oh}) consist of similar ingredients.
It is natural to assume that $J_{pd}$ in the phenomenological
model stems from the hybridization. Then it may be derived in the
Anderson-type model by means of the usual Schrieffer-Wolff
transformation, so that $J_{pd}\sim V^2/(\varepsilon_F - E_d)$.
Looking at Eq. (\ref{Ehol}) one can easily discern corresponding
contributions, which in fact were proposed for dilute magnets four
decades ago. \cite{AG} The second fraction in Eq. (\ref{Oh}) is
proportional to $m^*k_F$, i.e. to the Fermi momentum of holes in
the valence band. {\it Unlike} the phenomenological model, which
deals with localized spins and free holes near the top of the
valence band, our model takes into account the change in the
density of hole states (and therefore in their magnetic response)
due to resonance scattering and impurity band formation. One of
the results of this change is a more complicated dependence of the
magnetic coupling on the Mn content $x$ than the linear law
predicted by Eq. (\ref{Oh}). Another improvement of the mean-field
theory taken into account in Eq. (\ref{Curie}) is a more refined
description of the impurity magnetic moment. The actual moment $J$
arises as a vector sum of the moments of the d-electrons and the
bound p-hole, One should emphasize that at a high enough
concentration all mean-field descriptions fail because the alloy
approaches the instability region and the Mn distribution becomes
strongly inhomogeneous. In the case of Ga$_{1-x}$Mn$_x$As this is,
apparently, the region of $x>0.07$.

Next, one should discuss the relation between our approach and the
so called "ab initio" numerical calculations by means of the
local-density functional method.\cite{sanvito,Sato} Doping by Mn
atoms is modelled in these calculations by means of a finite-size
cluster of a GaAs or GaN host with one or several atoms replaced
by Mn. Then periodic structures are constructed from magnetic
clusters ("supercells") and the electronic bands in these
artificial objects are calculated in the local-spin-density
approximations (LSDA). So, the starting point in our approach and
the LSDA calculations is the same. No wonder that the positions of
CFR and DBH levels in these calculations are in good agreement
with those used in our model, based on previous single impurity
calculations.\cite{Kikoin,Zunger}  However, a direct comparison of
the two procedures, as far as the magnetic properties are
concerned, is somewhat problematic. The self-consistent LSDA
method results {\em by construction} only in a Stoner-like
itinerant magnetism. Therefore, magnetic states are discussed in
Refs. \onlinecite{sanvito,Sato} in terms of exchange splitting and
majority/minority spin subbands. It is difficult to discern the
genuine kinematic double exchange in this type of calculations.

To overcome this difficulty, Sanvito {\em et al}~ \cite{sanvito}
tried to fit their LDSA calculation by a free-electron model with
an effective hole mass $m^*$ and uniformly distributed impurities
described by a model square potential containing spin-independent
and exchange components ($W(r)$ and $J(r)$, respectively). Such
separation is purely phenomenological. It ignores the resonance
origin of the exchange potential. Besides, short-range impurity
scattering cannot be described in the effective mass
approximation.\cite{Fleurov2}  Nevertheless, the estimate of the
magnetic component $|J|=1.05$ eV is in a good agreement with the
corresponding parameter of our theory $\bar{z}V^2/E_{CFR}=1.034$
eV for the parameters used in the fitting of Fig. \ref{f.3} (the
value of $W=0.027$ eV is irrelevant because of the above mentioned
effective mass approximation). One should emphasize, however, that
the double-exchange coupling in our model is determined not by
this parameter, but by the integral $\Delta E_{FM}$ (\ref{Ehol}),
and this difference is in fact the benchmark for the discussion of
the differences between the indirect $pd$ exchange of
Vonsovskii-Zener type \cite{VZ} and the kinematic double
exchange.\cite{Zener}

A simplified picture of the valence band structure used in our
model (single heavy-hole band with semielliptic density of states)
allowed us to describe the dependence $T_C(x,p)$ using a minimal
number of fitting parameters. We have seen that a good
quantitative agreement between theory and experiment could be
achieved even with this very restricted set of parameters. With
the help of a more realistic energy band scheme (e.g. by taking
the light hole band into account), the restrictive symmetry of the
density of states would be removed, and the fitting procedure
would become more flexible. We intentionally imposed such severe
restrictions on the model to demonstrate its explanatory
capabilities. The above limitations will be lifted in future
studies.

Further progress in the theoretical description of ferromagnetism
in Mn-doped semiconductors is intimately related to the progress
in sample preparation and characterization. At present more or
less detailed data on $T_C(x,p)$ in {\em{p}}-type (Ga,Mn)As are
available, and we succeeded in a quantitative description of these
data within our model (Section III). We expect that the same
approach is applicable to {\em{p}}-type (Ga,Mn)P, but the scanty
experimental data do not allow us to check this expectation. As
for (Ga,Mn)N, the accumulation of experimental data is in
progress, and the most actual task is to reveal distinctions
between {\em{p}} and {\em{n}} type magnetic alloys. Recent
experimental data \cite{ngp} confirm existence of magnetic order
in $n$-(Ga,Mn)N, although the experimental state-of-the-art is far
from providing trustworthy $T_c(x)$ curves for theoretical
fitting. According to our theory, FM order is expected  both in
$p$- and $n$-type samples, but the latter case demands a serious
modification of the theory (Section IV) and a direct extrapolation
of the semi-empirical formula (\ref{Oh}) is questionable.

Another challenging question is the possibility of ferromagnetism
in elemental semiconductors and II-VI compounds doped by Mn and,
maybe, other magnetic ions and respective modification of the
theory. One can mention the recent reports about magnetism in Ge
doped by Mn \cite{Ge} and ZnO doped by Mn and Fe \cite{ZnO}. In
the latter case one deals with a wide-gap semiconductor, where
magnetic isoelectronic impurities Mn$^{2+}$ and Fe$^{2+}$ replace
Zn ions. The theory presented in Section IV seems to be applicable
in this case. Transition metal ions in Ge are as a rule
interstitial impurities sometimes involved in formation of complex
defects\cite{Kikoin,Zunger,Mod}, so the existing theoretical
approaches should be modified for this case. The most important
conclusion from our study is that the electronic structure of an
isolated magnetic impurity in the host material is the key to
understanding its behavior in concentrated magnetic alloys.
\begin{acknowledgments}
This work was supported by the Flemish Science Foundation
(FWO-VI), the Belgian Inter-University Attraction Poles (IUAP),
the "Onderzoeksraad van de Universiteit Antwerpen" (GOA) and the
Israeli Physical Society. VF and KK are grateful to Max-Planck
Institute for Complex Systems (Dresden) for partial support and
hospitality. VAI is grateful to Grenoble High Magnetic Field
Laboratory (Centre National de la Recherche Scientifique and the
Max-Planck Gesellschaft: CNRS-MPI)(Grenoble) for support and
hospitality. Authors acknowledge useful and fruitful discussions
with J. Cibert, K.W. Edmonds, B. Gallagher,   H. Hori, Y.H. Jeong,
T. Jungwirth, I. Ya. Korenblit, C. Lacroix, A. MacDonald, H.
Mariette, S. Sonoda
\end{acknowledgments}

\appendix
\section{ ENERGY SPECTRUM OF TWO COUPLED IMPURITIES}

The system of four Dyson equations for the $d$-electron Green
functions $G_{dii^\prime}^\sigma $ ($i,i^\prime$) has the
following form within the Hartree approximation:
\begin{equation}
G_{dii'}^\sigma(\varepsilon) = g_{d\sigma}(\varepsilon)\left(
\delta_{ii'}+ \sum_{\bf pp'}\sum_{j=1,2}V_{{\bf p}d}V_{{\bf
p'}d}^*e^{-i{\bf p}\cdot{\bf R}_i}G_{\bf pp'}^\sigma e^{i{\bf
p'}\cdot{\bf R}_l}G_{dji'}^\sigma \right). \label{AAA1}
\end{equation}
Here
\begin{equation}\label{dgreen}
g_{d\sigma}(\varepsilon) \delta_{ij} = (\varepsilon - E_d -
Un_i^{-\sigma} + i\delta \mbox{sign}(\varepsilon - \mu))^{-1}
\delta_{ij}
\end{equation}
is the bare single site d-electron Green function for the
$t_2\sigma$ electron centered at the ion $i$, taken in the Hartree
approximation. $\mu$ is the chemical potential. The Green function
for the band electrons can be found from the Dyson equation (a
similar system of equations for two Anderson impurities in metal
was analyzed in Ref. \onlinecite{Caroli})
\begin{equation}
\overline G_{\bf pp'}^\sigma(\varepsilon) = G_{\bf
pp'}^\sigma(\varepsilon) + \sum_{{\bf p"p'''}ij} G_{\bf
pp"}^\sigma(\varepsilon) V_{{\bf p"}di} g_{d\sigma}(\varepsilon)
V_{{\bf p^{'''}}dj}^* \overline G_{\bf p^{'''} p'}
^\sigma(\varepsilon)e^{i({\bf p"}\cdot{\bf R}_i - {\bf
p'''}\cdot{\bf R}_j)}.
\end{equation}
There are also impurity off-diagonal Green functions, which read
\begin{equation}\label{AAA2}
G_{di{\bf p}}^\sigma(\varepsilon) = g_{i\sigma}(\varepsilon)
V_{{\bf p^{'''}}d}^* \overline G_{\bf pp'}^\sigma(\varepsilon).
\end{equation}
The Green function $G_{\bf pp'}^\sigma$ describes the spectrum of
the valence band electrons modified by the two-impurity
short-range potential scattering. It satisfies the Dyson equation
\begin{equation}\label{AAA3}
G_{\bf pp'}^\sigma(\varepsilon) = g_{{\bf
p},\sigma}(\varepsilon)\delta_{\bf pp'} + g_{{\bf
p},\sigma}(\varepsilon) \sum_{\bf p"} W_{\bf pp"} G_{\bf
p"p'}^\sigma(\varepsilon)
\end{equation}
where
\begin{equation}\label{bgreen}
g_{{\bf p},\sigma}(\varepsilon) = (\varepsilon - \varepsilon_{\bf
p} + i\delta\mbox{sign}(\varepsilon - \mu))^{-1}.
\end{equation}

All the above equations give linear relations between the Green
functions. We may rewrite equation (\ref{AAA3}) using shorthand
notations for the Green function matrices
\begin{equation}\label{AAA4}
\textsf{G}_b = \textsf{A} \cdot \textsf{g}_b
\end{equation}
where $\textsf{g}_b$ stands for the diagonal matrix
(\ref{bgreen}). Equations (\ref{AAA1}) and (\ref{AAA2}) take the
form
\begin{equation}\label{AAA5}
\textsf{G} = \textsf{B} \cdot \textsf{g}_{0}
\end{equation}
where
\[
\sf{g}_{0} = \left(
\begin{array}{cc}
\sf{G}_b & 0\\
0 & \sf{g}_d
\end{array}
\right).
\]
with $\textsf{g}_d$ being a 2$\times$2 matrix (\ref{dgreen}). The
explicit expressions for the matrices \textsf{A} and \textsf{B}
can be readily found from the Dyson equations (\ref{AAA1}),
(\ref{AAA2}), and (\ref{AAA3}).

For further calculations we need determinants of the matrices
\textsf{A} and \textsf{B},
\begin{equation}\label{detA}
\textsf{Q}(\varepsilon) \equiv \det \textsf{A} = \textsf{q}
(\varepsilon)^2- W^2 L_{12}(\varepsilon) L_{21}( \varepsilon),
\end{equation}
where
$$\textsf{q}(\varepsilon) = 1-WL_{11}(\varepsilon)$$
and
\begin{equation}\label{sec1}
L_{ij} =\sum_{\bf p}g_p^0e^{i{\bf p}\cdot ({\bf R}_i-{\bf R}_j)}
\label{AGA}
\end{equation}
is the lattice Green function. We assume that $L_{11}=L_{22}$. The
second determinant is
\begin{equation}\label{detB}
\textsf{R} \equiv g_d^{-2}(\varepsilon) \det \textsf{B} =
[g_d^{-1}(\varepsilon) - V^2 M^\sigma_{11}( \varepsilon)]^2 - V^4
M^\sigma_{12}(\varepsilon) M^\sigma_{21}(\varepsilon).
\end{equation}
Here
\begin{equation}
M_{11}^\sigma  =  \sum_{\bf pp'}e^{-i({\bf p-p'})\cdot{\bf R}_1}
G_{\bf pp'}^\sigma = L_{11} + W \textsf{Q}^{-1}[L_{11}^2
\textsf{q} + L_{12}L_{21} \textsf{q} + 2 W L_{11}L_{12}L_{21}]
.\label{M11}
\end{equation}
$M_{22}^\sigma$ is obtained from Eq. (\ref{M11}) by exchanging
indices 1 and 2, and
\begin{equation}\label{M12}
M_{12}^\sigma  =  \sum_{\bf pp'}e^{-i{\bf p}\cdot{\bf R}_1 + i{\bf
p}'\cdot{\bf R}_2} G_{\bf pp'}^\sigma = L_{12} + W \textsf{Q}^{-1}
[L_{12} L_{11} \textsf{q} + L_{21} L_{11} \textsf{q} + WL_{21}^2
L_{12} ].
\end{equation}

Till now the calculations were made neglecting the orbital
degeneracy of the impurity $d$ states and within the Hartree
approximation. We can generalize these calculations using the
Hubbard I approximation for the $d$-electron Green functions (see,
e.g., Ref. \onlinecite{Kikoin3}). We assume also that the
three-fold degeneracy of the impurity $t_{2g}$ is not lifted and
the relevant physical quantities do not depend on the index $\mu$
enumerating these three states. The algebraic structure of the
Dyson equation is still the same as in (\ref{AAA1}). Then
considering the interaction of two three-fold degenerate states
belonging to the two impurities. This leads to the $6\times 6$
matrix
\begin{equation}\label{matrix} \textsf{B'}_{i\mu,j\mu'} =
\left(\begin{array}{cccccc}
      a & 0 & 0 & b & b & b \\
      0 & a & 0 & b & b & b \\
      0 & 0 & a & b & b & b \\
      b & b & b & a & 0 & 0 \\
      b & b & b & 0 & a & 0 \\
      b & b & b & 0 & 0 & a
    \end{array}\right)
\end{equation}
where
$$a = g_d^{-1}(\varepsilon) - V^2 K M^\sigma_{11}( \varepsilon),\ \
\ b = V^2 K M^\sigma_{12}(\varepsilon)$$
and $K = n_{d^5} + n_{d^4}$. $n_{d^5}$ is the probability that the
impurity $d$-shell is in the nondegenerate $d^5$ state, whereas
$n_{d^4}$ is the probability that the impurity d-shell is in one
of the three degenerate $d^4$ states. Calculating now the
determinant of the matrix (\ref{matrix}) one gets the equation
\begin{equation}\label{detB1}
\textsf{R} \equiv a^{-4} \det \textsf{B'} = [g_d^{-1}(\varepsilon)
- V^2 K M^\sigma_{11}( \varepsilon)]^2 - 9 V^4 K^2
M^\sigma_{12}(\varepsilon) M^\sigma_{21}(\varepsilon).
\end{equation}
which should be used instead of Eq. (\ref{detB}).

The occupation numbers $n_{d5}$ and $n_{d4}$ for the Hubbard-like
states obey a non-Fermi statistics, whose specific form in the
case considered here is
\begin{eqnarray}\label{occup}
n_{d5} = \frac{f(E_{CFR} - \mu)}{3 -2 f(E_{CFR} - \mu)}, \ \ \
n_{d4} = \frac{1 -  f(E_{CFR} - \mu)}{3 -2 f(E_{CFR} - \mu)}
\end{eqnarray}
If the chemical potential lies below the impurity level $E_{CFR}$
of the fifth electron in the d-shell then the Fermi distribution
$f(E_{CFR} - \mu)$ is zero at low temperatures and $n_{d5} =0,\
n_{d4} =1/3$, meaning that $K = 1/3$. If $E_{CFR} < \mu$ then
$n_{d5} =1,\ n_{d4} =0$,  and $K = 1$.

The energy (\ref{energy1}) can be found using the general property
of the Green functions, connecting their trace and determinant
\begin{equation}\label{property}
\mbox{Tr}~  \textsf{G}(\varepsilon) = \frac{d}{d \varepsilon}\ln
\det \textsf{G}(\varepsilon)
\end{equation}
Then using Eqs. (\ref{detA}) and (\ref{detB}), (\ref{property})
allows one to rewrite Eq. (\ref{energy1}) in the following form
\begin{equation}\label{calc3}
\Delta E = {\rm Im}\int_{-\infty }^{\infty }\frac{d\varepsilon
}{2\pi }\varepsilon \frac{d}{d \varepsilon }\left[ \ln {\sf
R}^{\sigma }(\varepsilon ) +  \ln {\sf Q}^{\sigma }(\varepsilon
)\right] = - {\rm Im} \int_{-\infty }^{\infty }\frac{d\varepsilon
}{2\pi }\left[ \ln {\sf R}^{\sigma }(\varepsilon )+  \ln {\sf
Q}^{\sigma }(\varepsilon )\right].
\end{equation}

The functions {\sf R} and {\sf Q} depend on the combination
$\varepsilon - i\mbox{sign}(\varepsilon - \mu)$, hence the
integration contour in (\ref{calc3}) can be deformed in such a way
as to embrace the cut from the band states and all the poles due
to the localized levels with the energies below the chemical
potential $\mu$, i.e. occupied states. Then equation (\ref{calc3})
transforms into (\ref{calc6}).

Next we consider a property of Eq. (\ref{calc3}), which will
simplify the calculation of the energy and provides a better
intuition to the results. Our model includes all the levels
belonging to the valence band with the addition of the impurity
d-levels, which interact with the band levels. Let us assume that
at zero temperature the chemical potential lies higher than all
these levels, meaning that they are all occupied. Then the total
energy of the system is
\begin{equation}\label{calc1}
E_{tot} = \mbox{Tr} \widehat{H}.
\end{equation}
In order to calculate this trace, we may represent the operator
$\widehat{H}$ in matrix form using the noninteracting band
states with the addition of the atomic $d$-functions as the basis.
Then the hybridization matrix elements will appear only in the
off-diagonal positions of this matrix, which do not influence the
result. The conclusion is that the energy $E_{tot}$ does not
depend on the value of the hybridization.

The potential scattering, $W_{\bf pp}$, contributes to the
diagonal elements of the Hamiltonian and may influence the value
of $E_{tot}$. However, we are interested here only in the indirect
exchange between the impurities. It can be found if we consider
the energy $\Delta E(R_{ij})$ and subtract from it the energy
corresponding to two noninteracting impurities,
\begin{equation}\label{calc8}
\Delta E_{ex} = - \frac{1}{\pi}{\rm Im}
\int_{\varepsilon_{hb}}^\mu d\varepsilon \left[ \ln \frac{{\sf
R}^{\sigma }(\varepsilon )}{{\sf R}_0^{\sigma }(\varepsilon )}+
\ln \frac{{\sf Q}^{\sigma }(\varepsilon )}{{\sf Q}^{\sigma
}_0(\varepsilon )}\right] + \Delta E_{loc}(\varepsilon < \mu).
\end{equation}
where ${\sf Q}^\sigma_0(\varepsilon )$ and ${\sf R}_0^{\sigma
}(\varepsilon )$ are obtained from (\ref{detA}) and (\ref{detB})
under the assumption that $L_{12}=L_{21}=0$. $\Delta
E_{loc}(\varepsilon < \mu)$ is the corresponding change of the
energies of the occupied localized levels. Now we may conclude
that the sum (\ref{calc8}) over all occupied states is equal to
the same sum over all empty states, but with the opposite sign.
Hence,
\begin{equation}\label{calc2}
\Delta E_{ex} = \frac{1}{\pi}{\rm Im} \int^{\varepsilon_{ht}}_\mu
d\varepsilon \left[ \ln \frac{{\sf R}^{\sigma }(\varepsilon
)}{{\sf R}_0^{\sigma }(\varepsilon )}+ \ln \frac{{\sf Q}^{\sigma
}(\varepsilon )}{{\sf Q}_0^{\sigma }(\varepsilon )}\right] -
\Delta E_{loc}(\varepsilon > \mu).
\end{equation}
Here $\Delta E_{loc}(\varepsilon > \mu)$ includes the empty
localized levels (if any) lying above the chemical potential.
These levels appear due to the combined action of both potential
($W$) and resonance ($V$) scattering mechanisms. Formally they can
be found as zeros of the determinant $\textsf{R}$ (\ref{detB}).

To find the contribution of the localized states to the magnetic
energy one should simply calculate the level positions modified by
the effective inter-impurity exchange due to hopping via empty
states and their occupation. We consider here two limiting cases.

First, we estimate the contribution of CFR levels, if they happen
to lie within the forbidden energy gap. It means that we may
expand the function $\textsf{R}(\varepsilon)$ close to the energy
$E_{CFR\sigma}^0$ of the isolated CFR level determined by the
equation
\begin{equation}
E_{CFR}^0=E_d + K V^2 P_{11}\left( E^0_{CFR}\right), \label{deep1}
\end{equation}
which describes the TM $d$-levels renormalized by their
hybridization with the {\em hh} band. Indirect inter-impurity
hopping results in splitting of two-impurity states and a shift of
localized states relative to the $hh$ band. These levels lie in
the discrete part of the spectrum, where the imaginary part of the
Green function (\ref{green}) $\Gamma_{ij}=0$. Neglecting potential
scattering, we obtain the equation
\begin{equation}
\label{detmBa} \textsf{R}  = [\varepsilon-E_d - K V^2
P^\sigma_{11}( \varepsilon)] [\varepsilon -E_d - K V^2
P^\sigma_{22}(\varepsilon)] - 9 K^2 V^4 P^\sigma_{12}(\varepsilon)
P^\sigma_{21}(\varepsilon)=0.
\end{equation}
for the two-impurity poles in the energy gap. The solution of Eq.
(\ref{detmBa}) is looked for in the form $ E_{CFR} = E_{CFR}^0 +
\delta E_{CFR}$. Now we expand the function ${\sf R}(E^0_{CFR} +
\delta E_{CFR})$ up to the second order terms with respect to $
\delta E_{CFR}$ and arrive at Eq. (\ref{level1}).

Second, we consider the case when the DBH levels lie in the
forbidden energy gap. Then the potential scattering $W$ is the
leading cause of the creation of the deep level. The energy of an
isolated DBH level corresponds to a zero of the function
$$\textsf{q}(E_{DBH}^0) = 1 - W P_{11}(E_{DBH}^0) = 0.$$
Then we look for zeros of the function $\textsf{R}(\varepsilon)$
at the energy $E_{DBH} = E_{DBH}^0 + \delta E_{DBH}$. Accounting
for the fact that both functions $\textsf{q}$ and $\textsf{Q}$ are
small in the vicinity of the energy $E_{DBH}^0$ the equation ${\sf
R}=0$ can be approximately rewritten in the form
\begin{equation}\label{DBH}
\left\{\textsf{q}^2 - W^2 P_{12}^2 - \frac{K V^2}{\Delta E} W
[P_{11}^2 \textsf{q} + P_{12}^2 \textsf{q} + 2 W P_{11} P^2_{12} ]
\right \}^2 =
\end{equation}
$$\frac{9 K^2 V^4}{\Delta E^2} \left\{P_{12} + W [2 P_{12} P_{11}
\textsf{q} + WP^2_{12}]\right\}^2$$
with $\Delta E = E_{DBH}^0 - E_d - V^2 P_{11}$. All the functions
$P_{ij}$ are now calculated at $\varepsilon = E_{DBH}^0$. We first
neglect the r.h.s term in Eq. (\ref{DBH}) and solve the
quadratic equation
\begin{equation}\label{DBH1}
\textsf{q}^2 - \frac{K V^2}{\Delta E} W [P_{11}^2 + P_{12}^2]
\textsf{q} - W^2 P_{12}^2 - \frac{2 K V^2 W^2}{\Delta E} P_{11}
P^2_{12}  = 0.
\end{equation}
>From here we obtain Eq. (\ref{splitq}) for the energy shifts due
to hopping between the two degenerate DBH levels.

When both these levels are empty we obtain a contribution to the
kinematic exchange by summing these two energies, extracting from
them the part due to the hybridization with the impurity d-states,
and changing the sign in the hole representation,
\begin{equation}\label{DBH2}
\Delta E_{DBH, ex} =  \frac{K V^2 P_{12}^2}{\Delta E P_{11}'}\;.
\end{equation}
Accounting for the rhs of Eq. (\ref{DBH}) will result in
higher order corrections, which can be neglected.


\begin{thebibliography}{99}

\bibitem[{*}]{perm}  Till December 2002 in Departement
Natuurkunde, Universiteit Antwerpen

\bibitem[\#]{aut2}  Electronic address: fleurov@post.tau.ac.il

\bibitem{Hayashi} M. Ilegems, R. Dingle, and L.W. Rupp (Jr.), J.
Appl. Phys. {\bf 46}, 3059 (1975); D.G. Andrianov, V.V. Karataev,
G.V. Lazarev, Yu.B. Muravlev, and A.S. Savel'ev, Sov. Phys.
Semicond. {\bf 11}, 738 (1977); T. Hayashi, Y. Hashimoto, S.
Katsumoto, and Y. Iye, Appl. Phys. Lett. {\bf 78}, 1691 (2001).

\bibitem{fmatsukara}  F. Matsukura, H. Ohno, A. Shen, and Y.
Sugawara, Phys. Rev. B {\bf 57}, R2037 (1998).

\bibitem{kikkawa}  J. M. Kikkawa and A. D. Awschalom, Nature
(London) {\bf 397}, 139 (1998).

\bibitem{Ohno0}  H. Ohno, H. Munekata, T. Penney, S. von
Moln\'ar, and L. L. Chang, Phys. Rev. Lett. {\bf 68}, 2664 (1992).

\bibitem{OhnoS}  H. Ohno, J. Magn. Magn. Mater. {\bf 200}, 110
(1999); Science {\bf 281}, 951 (1998).

\bibitem{Gall2}  K. W. Edmonds, K. Y. Wang, R. P. Campion, A. C.
Neumann, N. R. S. Farley, B. L. Gallagher, and C. T. Foxon, Appl.
Phys. Lett. {\bf 81}, 4991 (2002).

\bibitem{kw02} I. Kuryliszyn, T. Wojtowicz, X. Liu, J. K.
Furdyna, W. Dobrowolski, J.-M. Broto, M. Goiran, O. Portugal, H.
Rakoto, B. I. Raquet, Acta Phys. Pol. A {\bf 102}, 659 (2002).

\bibitem{kp03} K. C. Ku, S. J. Potashnik, R. F. Wang, S. H. Chun,
P. Schiffer, and N. Samarth, M. J. Seong and A. Mascarenhas, E.
Johnston-Halperin, R. C. Myers, A. C. Gossard, and D. D. Awschalom
Appl. Phys. Lett. {\bf 82}, 2302 (2003); K.W. Edmonds, P.
Boguslawski, K.Y. Wang, R.P. Campion, N.R.S. Farley, B.L.
Gallagher, C.T. Foxon, M. Sawicki, T. Dietl, M.B. Nardelli, and J.
Bernholc cond-mat/0307140.

\bibitem{g03} B. Gallagher, CECAM Workshop: "Diluted Magnetic
Semiconductors", June 12-14 2003, Lyon.

\bibitem{sonoda}  S. Sonoda, S. Shimizu, T. Sasaki, Y. Yamamoto,
and H. Hori, J. Appl. Phys. {\bf 91}, 7911 (2002).

\bibitem{reed}  M. L. Reed, N. A. El-Masry, H. H. Stadelmaier, M.
K. Ritums, M. J. Reed, C. A. Parker, J. C. Roberts, and S. M.
Bedair, Appl. Phys. Lett. {\bf 79}, 3473 (2001). N. Theodoropolou,
A. F. Hebard, M. E. Overberg, S. N. G. Chu, and R. G. Wilson,
Appl. Phys. Lett. {\bf 78}, 3475 (2001); M. E. Overberg, C. R.
Abernathy, S. J. Pearton, N. A. Theodoropoulou, K. T. McCarthy,
and A. F. Hebard, Appl. Phys. Lett. {\bf 79}, 1312 (2001).

\bibitem{crooker}  S. A. Crooker, D. A. Tulchinsky, J. Levy, D. D.
Awschalom, R. Garcia, N. Samarth, Phys. Rev. Lett. {\bf 75}, 505
(1995).

\bibitem{Vinc}  D. P. Di Vincenzo, Science {\bf 270}, 255 (1995).

\bibitem{ohno-field}  H. Ohno, D. Chiba, F. Matsukura, T. Omiya,
E. Abe, T. Dietl, Y. Ohno, and K. Ohtani, Nature (London) {\bf
408}, 944 (2000).

\bibitem{oiwa}  A. Oiwa, Y. Mitsumori, R. Moriya, T. Slupinski,
and H. Munekata, Phys. Rev. Lett. {\bf 88}, 137202 (2002).

\bibitem{Tanaka}  M. Tanaka and Y. Higo, Phys. Rev. Lett. {\bf
87,} 026602 (2001).

\bibitem{konig}  J. Konig, J. Schlieman, T. Jungwirth, and A. H.
MacDonald, in "Electronic Structure and Magnetism of Complex
Materials" (Springer Verlag, Berlin, 2002).

\bibitem{ohno-nature}  H. Ohno, D. Chiba, F. Matsukara, T.
Omiyama, E. Abe, T. Dietl, Y. Ohno, and K. Ohtani, Nature (London)
{\bf 408}, 944 (2000).

\bibitem{kawakami}  R. K. Kawakami, E. Johnston-Halperin, L. F.
Chen, M. Hanson, N. Gu\'ebels, J.S. Speck, A. C. Gossard, and D.
D. Awschalom, Appl. Phys. Lett. {\bf 77}, 2379 (2000).

\bibitem{BLee}  B. Lee, T. Jungwirth, and A.H. MacDonald, Phys.
Rev. B {\bf 65}, 193311 (2002).

\bibitem{Shen}  H. Ohno, A. Shen, F. Matsukara, O. Oiwa, A. Endo,
S. Katsumoto, and Y. Iye, Appl. Phys. Lett. {\bf 69}, 363 (1996).

\bibitem{Domcf}  T. Dietl, H. Ohno, F. Matsukura, J. Cibert, and
D. Ferrand, Science {\bf 287}, 139 (1998).

\bibitem{dietl2}  T. Dietl, H. Ohno, and F. Matsukura, Phys. Rev.
B {\bf 63}, 195205 (2001).

\bibitem{dietl3}  T. Dietl, F. Matsukura, and H. Ohno, Phys. Rev.
B {\bf66}, 033203 (2002).

\bibitem{Bhatt}  M. Berciu and R. N. Bhatt, Phys. Rev. Lett.
{\bf87}, 107203, 2001.

\bibitem{Semenov}  Yu. G. Semenov and S.M. Ryabchenko, Low Temp.
Phys. {\bf26}, 886 (2000)  [Fizika Nizkikh Temperatur {\bf 26},
1197 (2000)].

\bibitem{Duga}  V. K. Dugaev, V.I. Litvinov, J. Barna\'{s}, and
M. Vieira, Phys. Rev. B {\bf67}, 033201 (2003).

\bibitem{Inoue}  J. Inoue, S. Nonoyama, and H. Itoh, Phys. Rev.
Lett. {\bf 85}, 4610 (2000); Physica E {\bf 10}, 170 (2001).

\bibitem{sanvito}  S. Sanvito, P. Ordej\'{o}n, and N.A. Hill,
Phys. Rev. B {\bf 63}, 165206 (2001).

\bibitem{Sato} K. Sato and H. Katayama-Yoshida, Jpn. J. Appl.
Phys. {\bf 40}, L485 (2001); P. Mahadevan and A. Zunger,
"Electronic structure and ferromagnetism of 3d transition metal
impurities in GaAs" (preprint); L. Kronik, M. Jain, and J. R.
Chelikowsky, Phys. Rev. B {\bf 66}, 041203(R) (2002); M. van
Schilfgaarde and O. N. Mryasov, Phys. Rev. B {\bf 63}, 233205
(2001); E. Kulatov, H. Nakayama, H. Mariette, H. Ohta, and Y. A.
Uspenskii, Phys. Rev. B {\bf 66}, 045203 (2002).

\bibitem{Fleurov1}  V. N. Fleurov and K. A. Kikoin, J. Phys. C:
Solid State Phys. {\bf 9}, 1673 (1976).

\bibitem{Haldane}  F. D. M. Haldane and P. W. Anderson, Phys.
Rev. B {\bf 13}, 2553 (1976).

\bibitem{Kikoin}  K. A. Kikoin and V. N. Fleurov, {\em
"Transition Metal Impurities in Semiconductors",} (World
Scientific, Singapore, 1994).

\bibitem{Zunger}  A. Zunger, in {\em Solid State Physics,} Ed. by
H. Ehrenreich and D. Turnbull (Academic, Orlando, 1986) Vol. {\bf
39}, p. 276.

\bibitem{KIP} P. M. Krstaji\'c, V. A. Ivanov, F. M. Peeters, V.
Fleurov, and K. Kikoin, Europhys. Lett. {\bf 61}, 235 (2003).

\bibitem{Zener} C. Zener, Phys. Rev. {\bf 82}, 403 (1951).

\bibitem{schneider}  J. Schneider, U. Kaufmann, W. Wilkening, M.
Baeumler, and F. K\"{o}hl, Phys. Rev. Lett. {\bf 59}, 240 (1987).

\bibitem{Linnarson}  M. Linnarsson, E. Janzen, B. Monemar, M.
Kleverman, and A. Thilderkvist, Phys. Rev. B {\bf 55}, 6938
(1997).

\bibitem{szczytko}  J. Szczytko, A. Twardowski, K. Swiatek, M.
Palczewska, M. Tanaka, and T. Hayashi, K. Ando, Phys. Rev. B {\bf
60}, 8304 (1999).

\bibitem{Nagai}  Y. Nagai, T. Kunimoto, K. Nagasaka, H. Nojiri,
M. Motokawa, F. Matsukura, T. Dietl, and H. Ohno, Jpn. J. Appl.
Phys. {\bf 40}, 6231 (2001).

\bibitem{Kikoin77} K. A. Kikoin and V. N. Fleurov, J. Phys. C:
Solid State Phys. {\bf 10}, 4295 (1977).

\bibitem{Hemstreet}  L. A. Hemstreet, Phys. Rev. B {\bf 22}, 4590
(1980).

\bibitem{Picoli}  G. Picoli, A. Chomette, and M. Lannoo, Phys.
Rev. B {\bf 30}, 7138 (1984).

\bibitem{Singh}  V. A. Singh and A. Zunger, Phys. Rev. B {\bf
31}, 3729 (1985).

\bibitem{Fleurov2} V. N. Fleurov and K. A. Kikoin, J. Phys. C:
Solid State Phys. {\bf 19}, 887 (1986).

\bibitem{Kikoin2} K. A. Kikoin and V. N. Fleurov, J. Phys. C:
Solid State Phys. {\bf 17}, 2357 (1984).

\bibitem{graf}  T. Graf, M. Gjukic, L. G\"{o}rgens, O. Ambacher,
M. S. Brandt, and M. Stutzmann, Appl. Phys. Lett. {\bf 81}, 5159
(2002).

\bibitem{Anderson}  P. W. Anderson, Phys. Rev. {\bf 124}, 41
(1961).

\bibitem{Alex}  S. Alexander and P. W. Anderson, Phys. Rev. {\bf
133}, A1594 (1964).

\bibitem{Caroli} B. Caroli, J. Phys. Chem. Solids {\bf 28}, 1427
(1967).

\bibitem{Kikoin3} K. A. Kikoin and V. N. Fleurov, Zh. Ekps. Teor.
Fiz. {\bf 77}, 1062 (1979) [JETP {\bf 50}, 535 (1979)].

\bibitem{Mod}  {\em Data in Science and Technology -
Semiconductors: Group IV Elements and III-V Compounds,} edited by
O. Madelung (Springer-Verlag, Berlin, 1991).

\bibitem{Gall}  K. W. Edmonds, K. Y. Wang, R. P. Campion, A. C.
Neumann, C. T. Foxon, B. L. Gallagher, and P. C. Main, Appl. Phys.
Lett. {\bf 81}, 3010 (2002).

\bibitem{hebard} N. Theodoropoulou, A. F. Hebard, M. E. Overberg,
C. R. Abernathy, S. J. Pearton, S. N. G. Chu, and R. G. Wilson,
Phys. Rev. Lett. {\bf89}, 107203 (2002).

\bibitem{Barz} V. Barzykin, arXiv:cond-mat/0311114.

\bibitem{Asklund} H. Asklund, L. Ilver, J. Kanski, J. Sadowski,
and R. Mathieu Phys. Rev. B {\bf 66}, 115319 (2002).

\bibitem{Ohno2}  H. Ohno and F. Matsukura, Solid State Commun.
{\bf 117}, 179 (2001).

\bibitem{Yu}  K. M. Yu, W. Walukiewicz, T. Wojtowicz, W. L. Lim,
X. Liu, Y. Sasaki, M. Dobrowolska, and J. K. Furdyna, Appl. Phys.
Lett. \textbf{81}, 5 (2002).

\bibitem{Jung}  T. Jungwirth, J. K\"{o}nig, J. Sinova, J. Kucera,
and A.H. MacDonald, Phys. Rev. B {\bf 66}, 012402 (2002).

\bibitem{ks72} I. Ya. Korenblit and E. F. Shender, Zh. Exp. Teor.
Fiz. {\bf 62}, 1949 (1972) [Sov. Phys. - JETP {\bf 35}, 1017]; S.
L. Ginzburg, I. Ya. Korenblit, I. Y. Shender, Zh. Exp. Teor. Fiz.
{\bf 64}, 2255 (1973) [Sov. Phys. JETP {\bf 37}, 1141].

\bibitem{ks78} I. Ya. Korenblit and E.F. Shender, Usp. Fiz. Nauk
{\bf 126}, 233 (1978) [Sov. Phys. - Uspekhi {\bf 21}, 832 (1978)].

\bibitem{kss73} I. Ya. Korenblit, E. F. Shender, and B. I.
Shklovskii, Phys. Lett. {\bf 46}A, 275 (1973).

\bibitem{coval} P. W. Anderson, Phys. Rev. {\bf 79}, 350 (1950);
J. B. Goodenough, {\em Magnetism and Chemical Bond} (Interscience
Publ., N. Y., 1963).

\bibitem{McDo} T. Jungwirth, W. A. Atkinson, B. H. Lee, and A. H.
MacDonald, Phys. Rev. B {\bf59}, 9818 (1999).

\bibitem{AG} A.A. Abrikosov and L.P. Gorkov, Zh. Eksp. Teor. Fiz.
{\bf43}, 2230 (1962) [Sov. Phys. - JETP, {\bf16}, 1575 (1963)].

\bibitem{VZ} S. V. Vonsovskii, Zh. Exp. Teor. Fiz.
{\bf16}, 981 (1946); C. Zener, Phys. Rev. {\bf 81}, 440
(1951).

\bibitem{ngp} P.P. Chen, H. Makino, J.J. Kim, and T. Chao, J.
Cryst. Growth {\bf 251}, 331 (2003); R. Giraud, S. Kuroda, S.
Marcet, E. Bellet-Amarlic, X. Biquard, B. Barbara, D. Fruchart, J.
Cibert, and H. Mariette, Europhys. Lett., submitted

\bibitem{Ge} S. Cho, S. Choi, S.-C. Hong, Y. Kim, J. B. Ketterson,
B-J. Kim, Y. C. Kim, and J-H. Jung, Phys. Rev. B {\bf66}, 033303
(2002); A. Stroppa, S. Picozzi, and A. Continenza, A. J. Freeman,
PRB {\bf 68}, 155203 (2003)]; Y. D. Park, A. T. Hanbicki, S. C.
Erwin, C. S. Hellberg, J. M. Sullivan, J. E. Mattson, T. F.
Ambrose, and A. Wilson, Science {\bf 295}, 651 (2002); F. Tsui, L.
He, L. Ma, A. Tkachuk, Y. S. Chu, K. Nakajima, and T. Chikyow,
Phys.Rev.Lett. {\bf 91}, 177203 (2003).

\bibitem{ZnO} K. Ueda, H. Tabata, and T. Kawai, Appl. Phys. Lett.
{\bf79}, 988 (2001); D. P. Norton, S. J. Pearton, A. F. Hebard, N.
Theodoropoulou, L. A. Boatner, R. G. Wilson, Appl. Phys. Lett.
{\bf 82}, 239 (2003).

\end{thebibliography}
\end{document}